\documentclass[10pt,conference,letterpaper]{IEEEtran}
\IEEEoverridecommandlockouts
\usepackage{cite}
\usepackage{amsmath,amssymb,amsfonts}
\usepackage{algorithmicx}
\usepackage{graphicx}
\usepackage{textcomp}
\usepackage{xcolor}
\usepackage{array}
\usepackage{cases}

\usepackage[ruled,vlined]{algorithm2e}
\usepackage{ragged2e}
\usepackage{xpatch}
\makeatletter
\xpatchcmd\SetKwInOut
  {\hangafter=1\parbox[t]}
  {\hangafter=1\justify\parbox[t]}
  {}{\fail}
\SetKwInOut{Input}{Input}
\SetKwInOut{Output}{Output}
\usepackage[noend]{algpseudocode}

\usepackage[hyphens]{url}
\usepackage{breakurl}
\usepackage{multirow}
\usepackage{makecell}

\def\BibTeX{{\rm B\kern-.05em{\sc i\kern-.025em b}\kern-.08em
    T\kern-.1667em\lower.7ex\hbox{E}\kern-.125emX}}
\begin{document}

\title{EdgeTimer: Adaptive Multi-Timescale Scheduling in Mobile Edge Computing with Deep Reinforcement Learning}

\author{
	\IEEEauthorblockN{
		Yijun Hao\IEEEauthorrefmark{2},
		Shusen Yang\IEEEauthorrefmark{1},
		Fang Li\IEEEauthorrefmark{2},
		Yifan Zhang\IEEEauthorrefmark{2},
		Shibo Wang\IEEEauthorrefmark{2},
		and Xuebin Ren\IEEEauthorrefmark{2}}
	\IEEEauthorblockA{
		{${^\dagger}$}{${^\ast}$}Xi'an Jiaotong University, China \\
		Email: yijunhao@stu.xjtu.edu.cn, shusenyang@mail.xjtu.edu.cn, lifang1999@stu.xjtu.edu.cn, \\john\_efan@stu.xjtu.edu.cn, wshb20081996@stu.xjtu.edu.cn, xuebinren@mail.xjtu.edu.cn\\
		{${^\ast}$}Corresponding Author}
	}

\maketitle
\begin{abstract}
In mobile edge computing (MEC), resource scheduling is crucial to task requests' performance and service providers' cost, involving multi-layer heterogeneous scheduling decisions.
Existing schedulers typically adopt static timescales to regularly update scheduling decisions of each layer, without adaptive adjustment of timescales for different layers, resulting in potentially poor performance in practice.

We notice that the adaptive timescales would significantly improve the trade-off between the operation cost and delay performance.
Based on this insight, we propose EdgeTimer, the first work to automatically generate adaptive timescales to update multi-layer scheduling decisions using deep reinforcement learning (DRL).
First, EdgeTimer uses a three-layer hierarchical DRL framework to decouple the multi-layer decision-making task into a hierarchy of independent sub-tasks for improving learning efficiency.
Second, to cope with each sub-task, EdgeTimer adopts a safe multi-agent DRL algorithm for decentralized scheduling while ensuring system reliability.
We apply EdgeTimer to a wide range of Kubernetes scheduling rules, and evaluate it using production traces with different workload patterns.
Extensive trace-driven experiments demonstrate that EdgeTimer can learn adaptive timescales, irrespective of workload patterns and built-in scheduling rules. 
It obtains up to $9.1\times$ more profit than existing approaches without sacrificing the delay performance.
\end{abstract}


\section{Introduction}
\label{introduction}
Efficiently utilizing edge resources\textemdash which
are often more limited compared to cloud resources\textemdash matters for optimizing service performance, operation costs, and reliability in mobile edge computing (MEC) \cite{hu2015mobile, mao2017survey}.
Resource scheduling is a critical technique to achieve this.
It works by providing a set of scheduling decisions to assign available resources to task requests \emph{properly} \cite{jiang2019deep}. 
In the context of MEC, the resources of clouds and edges should be viewed as a unified whole that can be jointly scheduled, while they are commonly diversified in terms of capacity and duty \cite{ma2020cooperative}.
Efficient scheduling decisions are expected to exploit the collaboration of cloud servers, neighbor under-utilized edge servers, and the local edge server.
Typically, a three-layer heterogeneous architecture is presented by segregating the overall scheduling process into layers of scheduling decisions performed in different collaboration manners, including \emph{Layer-1: edge-cloud scheduling, Layer-2: edge-edge scheduling, and Layer-3: intra-edge scheduling} \cite{luo2021resource}.

The three layers of scheduling decisions \emph{are not executed in isolation}.
When processing a task request, there is a specific sequence and correlation involved in the decision-making process for different layers.
For example, edge-cloud service placement \cite{farhadi2019service}, edge-edge task offloading \cite{han2022edgetuner}, and intra-edge resource allocation \cite{xiong2020resource} successively determine which cloud services to pre-place in which edge servers before task arrivals, which servers to offload arrived tasks, and how many resources to allocate to offloaded tasks, respectively. 

\begin{figure*}[!t]
\begin{minipage}{0.46\linewidth} 
      \includegraphics[width=1\columnwidth]{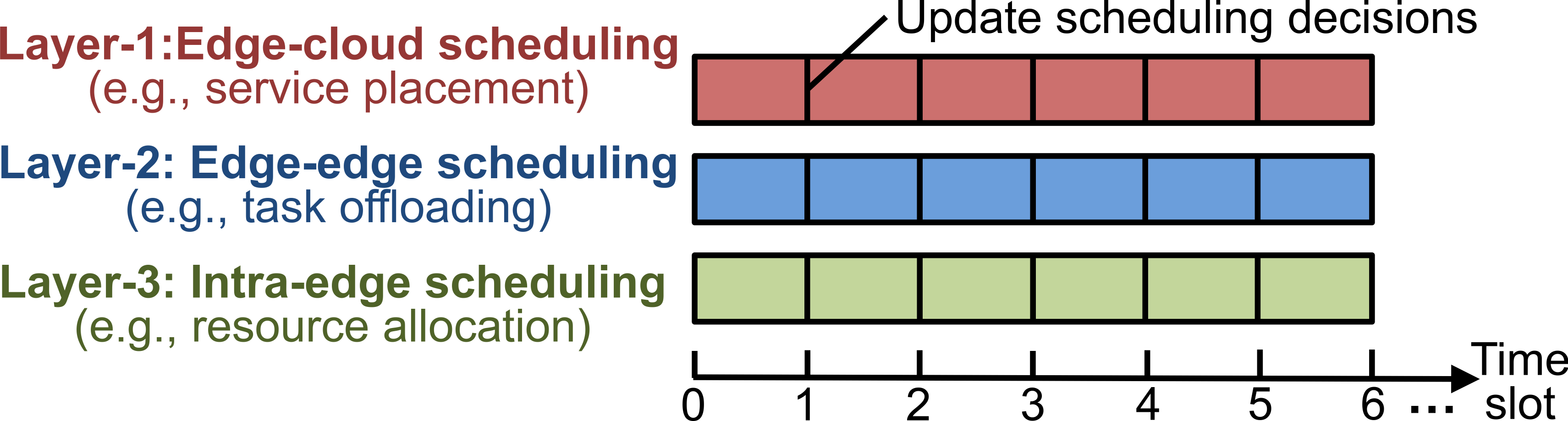} \\\footnotesize{(a) Static single-timescale scheduling.}
    \end{minipage}
    \begin{minipage}{0.26\linewidth} 
      \includegraphics[width=1\columnwidth]{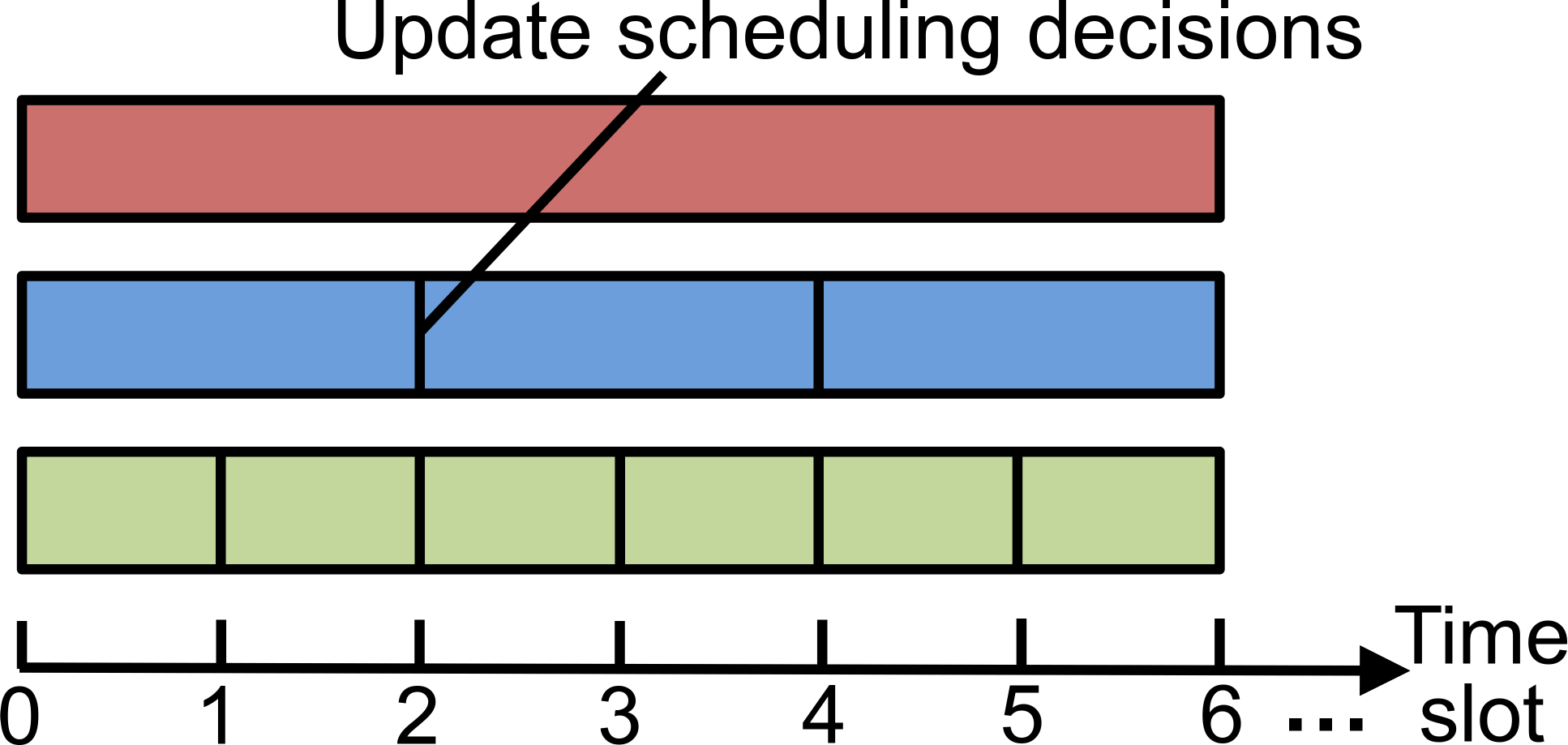} \\\footnotesize{(b) Static multi-timescale scheduling.}
    \end{minipage}
    \begin{minipage}{0.26\linewidth} 
      \includegraphics[width=1\columnwidth]{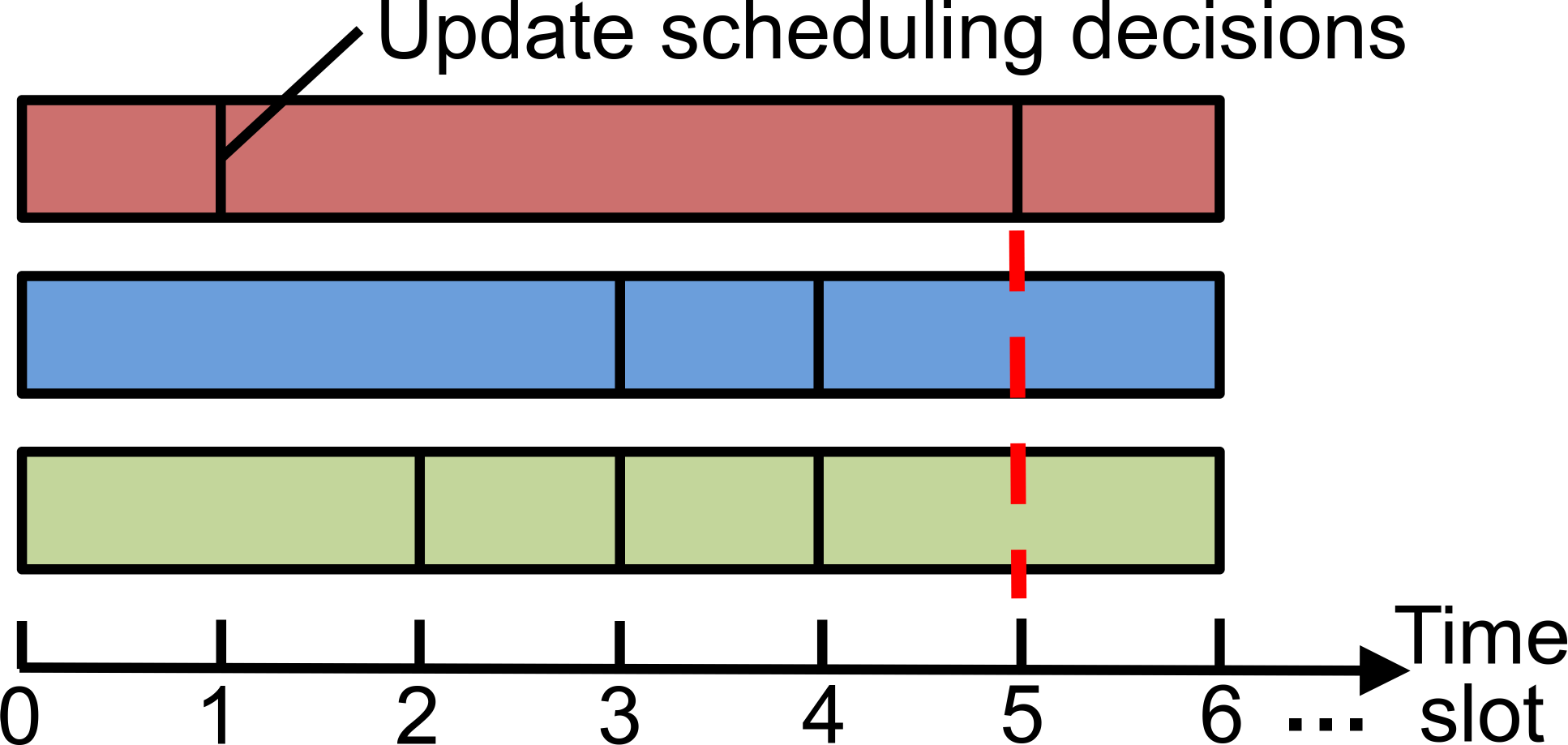} \\\footnotesize{(c) Adaptive multi-timescale scheduling.}
    \end{minipage}
\caption{Timescales of different resource scheduling frameworks in MEC.
The red dotted line shows the asynchrony nature of the adaptive multi-timescale framework, i.e., EdgeTimer, where the higher-layer scheduling decision is updated but lower-layer decisions remain unchanged.}
\vspace{-0.5em}
\label{intro}
\end{figure*}

Most MEC schedulers \cite{wang2022decentralized, zhang2021deepreserve, poularakis2019joint, ma2020cooperative} update scheduling decisions at a \textit{fixed} interval that is \textit{uniform} across all three layers, which we call \textit{static single-timescale mode} (see Fig. \ref{intro}(a)).
Typically, the updating interval is extremely short (e.g., one second) to ensure a timely decision adjustment based on real-time requests.
However, such frequent decision adjustment incurs high operation costs for service providers, such as costs of service migration \cite{guo2023efficient}, handover \cite{sun2017emm}, and resource reconfiguration \cite{jiao2017smoothed}.

Fewer schedulers \cite{farhadi2019service, han2021tailored} set up different updating intervals across multiple scheduling layers, but the intervals are still \textit{fixed}, which we call \textit{static multi-timescale mode} (see Fig. \ref{intro}(b)).
They typically use a \textit{constant} scaling ratio to determine the difference in timescales between adjacent scheduling layers, and consequently cannot adapt to environment variations.
For example, in Fig. \ref{intro}(b), updates in Layer-3 occur $2\times$ more frequently than in Layer-2 and $6\times$ more frequently than in Layer-1.
Furthermore, human-based tuning is heavily required to identify an appropriate scaling ratio that enables service providers to save costs while providing high-performance services.
It is a time-consuming and uneconomical process due to the additional labor costs involved.

That naturally raises a new critical question: \emph{How to automatically and dynamically determine respective timescales of multiple layers of scheduling decisions to achieve the trade-off between the operation cost and service performance?}
To solve this problem, we propose EdgeTimer, an \emph{adaptive multi-timescale} scheduling framework for edge-cloud, edge-edge, and intra-edge scheduling, as shown in Fig. \ref{intro}(c).
EdgeTimer has the following three desired features:

\textbf{F1: Adaptation to environment variations.} The timescale for each layer of scheduling decisions is time-varying, adapting to changeable patterns of online task requests.

\textbf{F2: Asynchrony over multi-layer scheduling.} It decouples multi-layer scheduling decisions, allowing lower-layer decisions to remain unchanged while higher-layer decisions are updated, as shown by the dotted red line in Fig. \ref{intro}(c).
The value of timescales is determined independently across different scheduling layers, instead of being strongly linked by a scaling ratio.

\textbf{F3: Autonomy for edge servers.} 
To reduce communication cost from the cloud to edges, e.g., high transmission delay, each edge server is required to determine its timescales \textit{locally}, without resorting to the computation of remote clouds.

EdgeTimer uses deep reinforcement learning (DRL) \cite{mnih2015human} to learn update policies, i.e., deciding whether to update current scheduling decisions of each layer for each edge server at each slot.
DRL is well-suited to our problem as it allows to learn update policies purely through the actual network conditions and task requests.
However, when applying off-the-shelf DRL to our problem, two main challenges need to be addressed. 
 
First, conventional DRL algorithms are designed to achieve a high-level aggregated goal.
In our problem, we define the overall goal as \emph{service providers' profit} that explicitly measure the cost-performance trade-off.
Correspondingly, the learning task is set to simultaneously learn three-layer policies to jointly maximize the expected profit. 
This makes the learning task complicated and intractable, as improper policies of any layers may lead to poor profit performance due to inter-layer dependencies.
We design a \emph{three-layer hierarchical DRL framework} to decompose the overall learning task into three independent hierarchical sub-tasks based on temporal abstraction, each of which is associated with an individual sub-goal and is handled by a DRL controller.
This approach significantly reduces the task complexity, and improves learning efficiency by mitigating the effect of any layers of improper policies.

Second, conventional DRL algorithms usually use one agent for centralized decision-making, and are incapable of decentralized scheduling tasks due to non-stationary environment issues \cite{lowe2017multi}.
To this end, for each layer of DRL controllers, we design a \emph{safe multi-agent DRL approach} to learn layer-specific decentralized policies.
To tackle the lack of global information, the non-local information from other agents is allowed to use for ease of offline training, while it is not observed in the online process.
Moreover, to ensure the reliability of online decisions, EdgeTimer also integrates a safe learning scheme to mask unsafe decisions, e.g., inter-procedure conflicting decisions, caused by uncertainty of neural networks.

We integrated EdgeTimer with a realistic Vienna system-level simulator \cite{muller2018flexible} and evaluated it using multi-pattern workload traces derived from Alibaba's production clusters \cite{Ali}.
A total of $45$ typical scheduling rules were implemented to act as built-in scheduling rules, including Kubernetes rules \cite{k8s} and other widely-used rules.
Extensive trace-driven experiments show that by applying EdgeTimer on $45$ typical scheduling rules, it improves the service providers' profit by up to $9.1\times$ over existing schedulers under the same rule, without sacrificing the delay performance of task requests. 

To our knowledge, EdgeTimer is \emph{the first to automatically and collaboratively control the update timescales of multiple scheduling decisions to optimize the profit.}
Service providers would benefit from our work in three ways:
1) They can directly improve their profit by providing high-performance services with no unnecessary costs.
2) EdgeTimer is not intrusive to current scheduling approaches of service providers.
It only provides an update signal when it is necessary.
Service providers can decide whether to rely on the signal based on their extensive experience.
3) Besides the application case in this paper, they can extend EdgeTimer to other single or multiple layers of scheduling decisions by simply changing the state and reward settings of DRL controllers.

\section{Motivating Example}
\label{motivation}
\begin{figure*}[!t]
\centerline{\includegraphics[width=2\columnwidth]{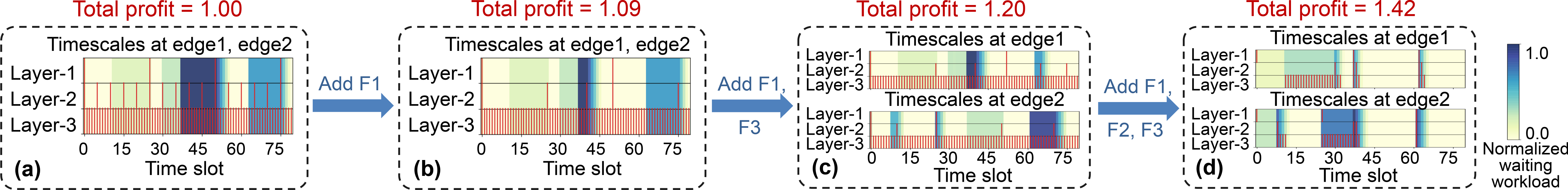}}
\caption{An example to show that by \emph{cumulatively} applying EdgeTimer's features of adaptation (F1), asynchrony (F2) and autonomy (F3) to existing approaches, the profit can be improved significantly.
The scheduling decisions are updated at the moment of the red vertical lines.
}
\vspace{-1em}
\label{motivation fig}
\end{figure*}
EdgeTimer introduces a new dimension (i.e., updating timescales) to the research issue of MEC scheduling, in order to further improve the trade-off between system performance and costs.
We used the \textit{profit} (detailed in Sec. \ref{profit discreb}) as a metric to judge the advancement of performance-cost trade-offs.
Figure \ref{motivation fig} provides an example to demonstrate that by adding the EdgeTimer's features, i.e., adaptation (\textbf{F1}), asynchrony (\textbf{F2}), and autonomy (\textbf{F3}), to the existing approaches step by step, the profit can be improved significantly.

We consider two edge servers and a cloud server in a MEC system.
Each of the edge servers performs three layers of scheduling decisions, including Layer-1: edge-cloud service placement, Layer-2: edge-edge task offloading, and  Layer-3: intra-edge resource allocation.
Figure \ref{motivation fig}(a) visualizes the updating timescales and scheduling results of state-of-the-art schedulers, where a fixed scaling ratio is used to determine the difference in updating timescales between adjacent scheduling layers.
For a fair comparison, we iterate through all possible scaling ratios in a given set to select the optimal one as the scaling ratio used in Fig. \ref{motivation fig}(a).
All edge servers adopt the same timescales.
The total profit obtained by the state-of-the-art scheduler is normalized to one.

In Fig. \ref{motivation fig}(b), we adjust the inter-layer scaling ratio based on the total waiting workloads of two edge servers, i.e, applying the adaptation feature of EdgeTimer (\textbf{F1}).
Fig. \ref{motivation fig}(b) shows that by employing a lower decision update frequency (the sparser red vertical lines), a lower waiting workload than Fig. \ref{motivation fig}(a) can be attained (the shorter dark purple blocks).
The adaptive updating timescales significantly reduce unnecessary operation costs, achieving $1.09\times$ higher profit than state-of-the-art schedulers.
On this basis, the profit can be improved if we follow the autonomy feature of EdgeTimer (\textbf{F3}).
Each edge server is allowed to use differentiated timescales according to local waiting workloads (Fig. \ref{motivation fig}(c)).
The total profit is $1.2\times$ higher than state-of-the-art schedulers.

Notably all the above practices rely on inter-layer scaling ratios to control three-layer scheduling simultaneously.
Suggested by the asynchrony feature of EdgeTimer (\textbf{F2}), the profit can be improved by updating three-layer scheduling decisions asynchronously.
We set the workload threshold for each layer and update the corresponding decisions only when the current workload exceeds the layer-specific threshold (Fig. \ref{motivation fig}(d)).
Finally, the approach that integrate all EdgeTimer's features makes $1.42\times$ higher profit than state-of-the-art schedulers.

Though this example demonstrates the potential of using EdgeTimer to improve the profit, its effectiveness highly relies on the prior knowledge.
Once the workload pattern changes, we have to tune parameters, e.g., workload threshold, for high performance.
In this paper, driven by DRL, EdgeTimer is able to automatically determine timescales, and as a result, it achieves much higher profits than all the above practices.

\section{System model and problem formulation}
\label{system model sub}
\begin{figure}[!t]
\centerline{\includegraphics[width=0.85\columnwidth]{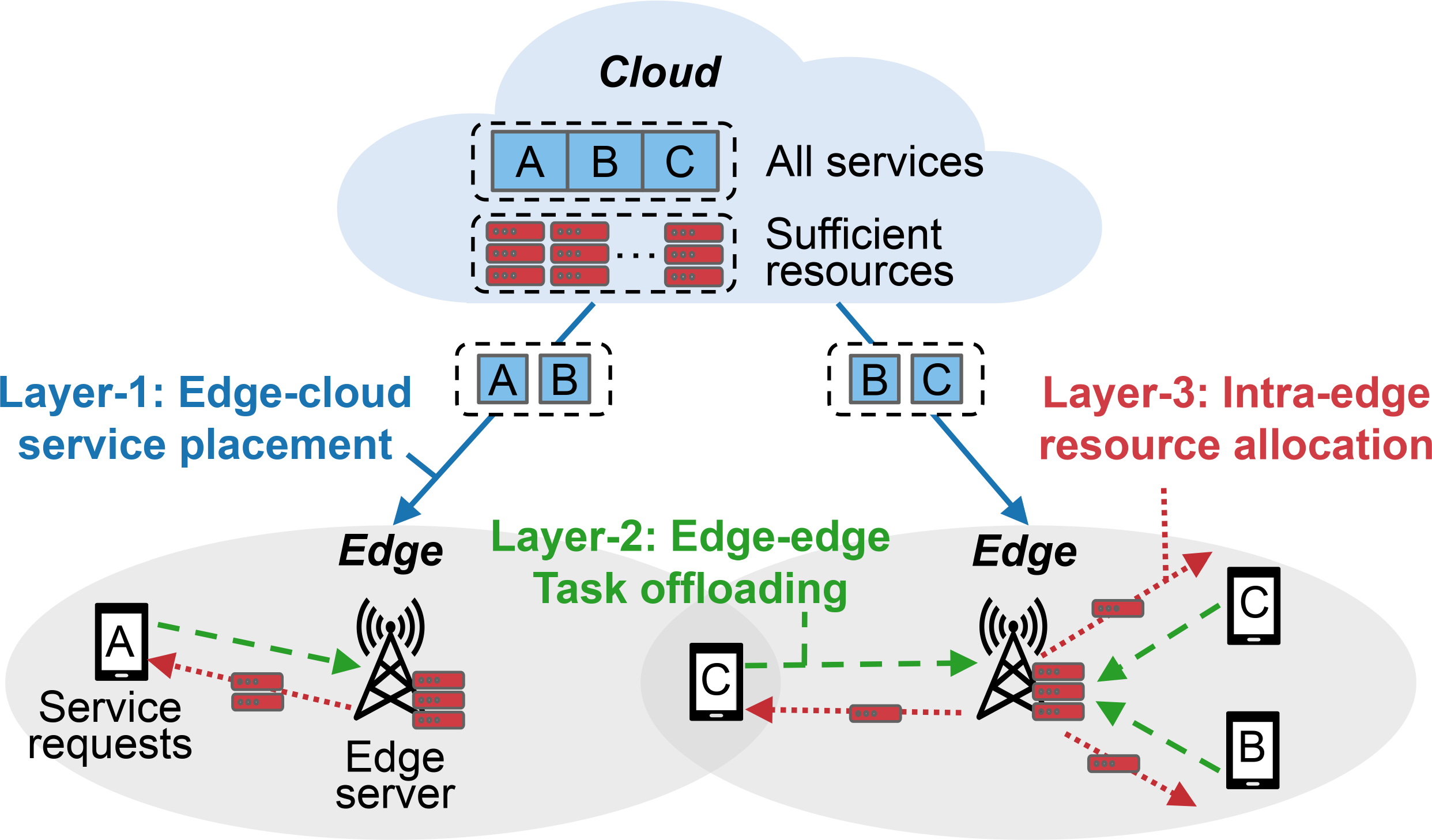}}
\caption{The illustration of three-layer heterogeneous scheduling in MEC.}
\vspace{-1em}
\label{system model}
\end{figure}
Our objective is to optimize the overall profit by updating scheduling decisions with an \textit{adaptive} timescale in heterogeneous MEC systems.

\subsection{MEC System Model}
As shown in Fig. \ref{system model}, we consider a MEC system proceeding in discrete time slots $\mathcal{T}=\{0,1,...,t,...\}$.
$N$ edge servers, $\mathcal{N}=\{1,...,N\}$, are located at respective base stations, each of which is equipped with limited computation, storage, and communication resources.
All edge servers are connected via back-haul links \cite{farhadi2019service}.
A cloud $o$ with sufficient resources is connected to edge servers by core networks \cite{ma2020cooperative}.

\subsection{Resource Scheduling Model}
At each time slot $t$, task requests arrive at edge servers stochastically.
Efficient processing of tasks relies on three-layer heterogeneous scheduling with different collaborations, including \emph{Layer-1: edge-cloud scheduling, Layer-2: edge-edge scheduling, and Layer-3: intra-edge scheduling} \cite{luo2021resource}.
We apply EdgeTimer to three typical cases corresponding to the above three layers, i.e., edge-cloud service placement \cite{farhadi2019service}, edge-edge task offloading \cite{han2022edgetuner}, and intra-edge resource allocation \cite{xiong2020resource}.

\emph{1) Layer-1: Edge-Cloud Service Placement:}
Each task request is associated with an application, e.g., machine learning inference or augmented reality. 
To process a task, the edge server must pre-cache the application-specific service, e.g., trained models and related databases \cite{poularakis2019joint}.
Different services occupy different storage spaces.
The cloud caches all services $\mathcal{S}=\{1,...,S\}$.
Considering storage limits of edge servers, only a subset of services can be placed on each of them at a given slot.
Service placement provides decisions $x_{i,s}(t)$ indicating whether service $s$ is placed on edge $i$ at slot $t$.

\emph{2) Layer-2: Edge-Edge Task Offloading:} 
To better utilize limited edge resources, tasks arrived at the local edge server $i$ can be offloaded to an under-utilized neighbor edge $j$ that has cached corresponding services.
Decisions of task offloading $y_{i,j,s}(t)$ determine whether a task of service $s$ is offloaded from edge server $i$ to server $j$ at slot $t$.
The offloading process consumes the communication resources between the original and destination servers, and incurs transmission delay.

\emph{3) Layer-3: Intra-Edge Resource Allocation:}
It focuses on how much computation resources are allocated by edge $i$ to offloaded service tasks $s$ at slot $t$, denoted by $z_{i,s}(t)$.
There may exist waiting queues in the buffer of edge servers.
Thus, the computation delay required for allocation decisions includes the queueing delay and execution delay.

Please note that the three-layer scheduling decisions, i.e., $x_{i,s}(t)$, $y_{i,j,s}(t)$ and $z_{i,s}(t)$, are \emph{tightly coupled.}
The higher-layer decisions determine lower-layer feasible regions, e.g., the tasks can only be offloaded to servers that have placed corresponding services.
In contrast, the lower-layer scheduling results reflect the effectiveness of higher-layer decisions.

\subsection{Profit Model}
\label{profit discreb}
The profit earned by service providers can be defined by subtracting their operation cost from their revenue.

\emph{1) Revenue:}
Service providers get revenue by selling computation resources to users \cite{huang2021platform,zang2021soar}.
Considering that delayed tasks have a negative impact on the revenue, we use a more realistic metric to represent the revenue obtained by server $i$ at slot $t$ as:
\begin{equation}
\setlength{\abovedisplayskip}{-1pt}
\setlength{\belowdisplayskip}{-1pt}
g_{i}(t) = \sum_{s\in\mathcal{S}}\mu_{i,s}(t)p_{s}(t),
\label{revenue}
\end{equation}
where $\mu_{i,s}(t)$ is the total amount of computation resources allocated to service tasks $s$ that were \emph{processed within delay budgets} by server $i$ at slot $t$.
Given that servers are allowed to differentiate the unit price of each service task for profit maximization \cite{Three2020}. 
Furthermore, the unit price usually decreases as the billing cycle increases in real markets \cite{microsoft}.
We denote the unit price of computation resources using a service-specific time-varying parameter $p_{s}(t)$.

\emph{2) Operation Cost:}
Too frequent adjustment of scheduling decisions, i.e., $x_{i,s}(t)$, $y_{i,j,s}(t)$, and $z_{i,s}(t)$, would increase operation cost.
Specifically, placing a service $s$ requires the edge server $i$ to download non-trivial data from server $j$ of a neighbor edge or a remote cloud, which incurs the operation cost, $C^{1}_{i,j,s}$, defined by the service type and transmission distance \cite{farhadi2019service}.
Following a general assumption \cite{farhadi2019service}, removing an installed service does not increase the cost.
The cost of placing a new service $s$ on edge $i$ is $C^{1}_{i,s}\!=\!\min_{j\in\{j|x_{j,s}(t-1)=1\}}\!C^{1}_{i,j,s}$.
The operation cost for service placement of edge $i$ at slot $t$ can be expressed as:
\begin{equation}
c^{1}_{i}(t) = \sum_{s\in\mathcal{S}}C^{1}_{i,s}\mathbb{I}_{\{1\}}(x_{i,s}(t)-x_{i,s}(t-1)),
\label{placement cost}
\end{equation}
where $\mathbb{I}_{\{1\}}(x)$ is an indicator function that has value of $1$ if $x\in\{1\}$ and $0$ otherwise.
Furthermore, if a task is offloaded from a current server $k$ to another server $m$, it incurs operation cost $C^{2}_{k,m,s}$ including data transmission cost and service interruption cost caused by the inter-server handover \cite{sun2017emm}.
The offloading cost of edge $i$ at slot $t$ is defined as:
\begin{equation}
c^{2}_{i}(t) = \sum_{s\in\mathcal{S}}C^{2}_{k,m,s}\mathbb{I}_{\{1\}}(y_{i,k,s}(t-1)y_{i,m,s}(t)).
\label{offloading cost}
\end{equation}
Changing resource allocation decisions may lead to resource reconfiguration, and incurs the resource lead time caused by the server booting and initialization \cite{jiao2017smoothed,wang2018moera}.
Since shutting down a server is much faster than booting it, we use the increase in resources allocated as the operation cost of resource reconfiguration.
Let $C^{3}_{i}$ be the one-unit reconfiguration cost of the computation resource.
At time slot $t$, the allocation cost of edge $i$ can be represented as:
\begin{equation}
c^{3}_{i}(t) = \sum_{s\in\mathcal{S}}C^{3}_{i}[z_{i,s}(t)-z_{i,s}(t-1)]^{+}.
\label{allocation cost}
\end{equation}

\subsection{Problem Formulation}
\label{problem formulation sub}
Our objective is to \emph{maximize the long-term profit} by jointly optimizing timescales that are used to update three-layer scheduling decisions. 
The control variables are $a^1_{i}(t)$, $a^2_{i}(t)$, and $a^3_{i}(t)$, indicating \emph{whether current placement, offloading, and allocation decisions at edge server $i$ need to be updated, respectively}.
Existing scheduling algorithms act as built-in rules to provide candidate scheduling decisions for placement $\widetilde{x}_{i,s}(t)$, offloading $\widetilde{y}_{i,j,s}(t)$, and allocation $\widetilde{z}_{i,s}(t)$.
For example, if $a^1_{i}(t)=1$, the server $i$ would perform placement decisions $\widetilde{x}_{i,s}(t)$ generated by existing placement rules.
Conversely, the server $i$ would continue previous decisions.
The placement decisions performed in edge $i$ at slot $t$ can be expressed as:
\begin{align}
\setlength{\abovedisplayskip}{-1pt}
&{x}_{i,s}(t)=\Gamma^{1}({a^1_{i}(t)})
  =\begin{cases}
    \widetilde{x}_{i,s}(t),&a^1_{i}(t)=1,\\
    \Gamma^{1}({a^1_{i}(t-1)}),&a^1_{i}(t)=0.
  \end{cases}
\end{align}
The joint optimization problem can be formulated as:
\begin{align}
\mathcal{\widetilde{P}}\!\!: \!&\max \mathbb{E}[\sum_{t=0}^{\infty}\!\sum_{i\in\mathcal{N}}(g_{i}(t)\!-\!a^1_{i}(t)c^{1}_{i}(t)\!-\!a^2_{i}(t)c^{2}_{i}(t)\!-\!a^3_{i}(t)c^{3}_{i}(t))]\nonumber\\
&\text { s.t. }\quad\quad a^1_{i}(t), a^2_{i}(t), a^3_{i}(t) \in\{0,1\}, \forall{t,i}.
\label{general problem}
\end{align}

\section{EdgeTimer Design}
\label{design}
In EdgeTimer, we design a three-layer HDRL framework to decouple the joint optimization problem. 
Then, the safe MADRL algorithm for each sub-problem is explained.
\subsection{Three-Layer Hierarchical DRL for Problem Decomposition}
Achieving the objective in (\ref{general problem}) relies on performing three layers of tightly coupled decision-making.
This makes it intractable to generate three effective decisions simultaneously.
For ease of learning, we decouple the original problem into three independent sub-problems, and design a three-layer hierarchical deep reinforcement learning (HDRL) framework to solve different sub-problems in a coordinated manner.

\emph{1) Hierarchical Decomposition for the Original Problem:}
We decouple the original problem ($\mathcal{\widetilde{P}}$) into sub-problems ($\mathcal{P}1,\mathcal{P}2,\mathcal{P}3$) by decomposing the total objective into respective sub-revenue and sub-cost of different layers of scheduling.

\textbf{Edge-cloud service placement sub-problem ($\mathcal{P}$1).}
The placement cost is derived from (\ref{placement cost}).
Although the revenue cannot be directly decomposed, the extent of service coverages can potentially measure the contribution of current placement decisions to the total revenue, as verified in subsection \ref{EdgeTimer Deep Dive}.
The service placement sub-problem can be formulated as:
\begin{align}
\setlength{\abovedisplayskip}{-1pt}
\setlength{\belowdisplayskip}{-1pt}
\mathcal{P}1: &\max \mathbb{E}[\sum_{t=0}^{\infty}\alpha d(t)-\sum_{t=0}^{\infty}\sum_{i\in\mathcal{N}}a^1_{i}(t)c^{1}_{i}(t)]\nonumber\\
&\text { s.t. }\quad\quad a^1_{i}(t)\in\{0,1\}, \forall{t,i},
\label{subproblem 1}
\end{align}
where $d(t)$ is the amount of \!computation demands of tasks covered by placed services at slot $t$.
It is associated with current placement decisions, i.e., $d(t)=d(\Gamma^{1}({a^1_{1}(t)}), ..., \Gamma^{1}({a^1_{N}(t)}))$.
The coefficient $\alpha$ is to weigh the impact of service coverages.

\textbf{Edge-edge task offloading sub-problem ($\mathcal{P}$2).}
Given the offloading sub-cost in (\ref{offloading cost}), a straight option to formulate the offloading sub-problem is to set the transmission delay budget and maximize the difference between total computation demands of current tasks, $u(t)$, that complete offloading within the transmission delay budget and the sub-cost.
However, the above option is not equivalent to maximizing the original profit, as it may result in offloading too many workloads to the nearest edges, inducing high computation delay and revenue loss. 
To avoid load imbalance, we introduce a load balancing function $v(\cdot)$ that maps current offloading decisions $\Gamma^{2}({a^2_{i}(t)})_{i\in{\mathcal{N}}}$ to the workload variance of all edge servers.
The task offloading sub-problem can be given by:
\begin{align}
\setlength{\abovedisplayskip}{-1pt}
\setlength{\belowdisplayskip}{-1pt}
\mathcal{P}2: &\max \mathbb{E}[\sum_{t=0}^{\infty}\beta u(t)-\sum_{t=0}^{\infty}\sigma v(t)-\sum_{t=0}^{\infty}\sum_{i\in\mathcal{N}}a^2_{i}(t)c^{2}_{i}(t)]\nonumber\\
&\text { s.t. }\quad\quad a^2_{i}(t)\in\{0,1\}, \forall{t,i},
\label{subproblem 2}
\end{align}
where $\beta$ and $\sigma$ are coefficients to control the degree of the offloaded task satisfaction and workload variance, respectively.

\textbf{Intra-edge resource allocation sub-problem ($\mathcal{P}$3).}
Similarly, we set computation delay budgets for service requests, and denote the total computation demands of current tasks processed within computation delay budgets and the corresponding weight coefficient as $l(t)$ and $\phi$, respectively.
The resource allocation sub-problem is defined by:
\begin{align}
\setlength{\abovedisplayskip}{-1pt}
\setlength{\belowdisplayskip}{-1pt}
\mathcal{P}3: &\max \mathbb{E}[\sum_{t=0}^{\infty}\phi l(t)-\sum_{t=0}^{\infty}\sum_{i\in\mathcal{N}}a^3_{i}(t)c^{3}_{i}(t)]\nonumber\\
&\text { s.t. }\quad\quad a^3_{i}(t)\in\{0,1\}, \forall{t,i}.
\label{subproblem 3}
\end{align}

\emph{2) Workflow of Three-Layer Hierarchical DRL Framework:}
\begin{figure}[!t]
\centerline{\includegraphics[width=0.95\columnwidth]{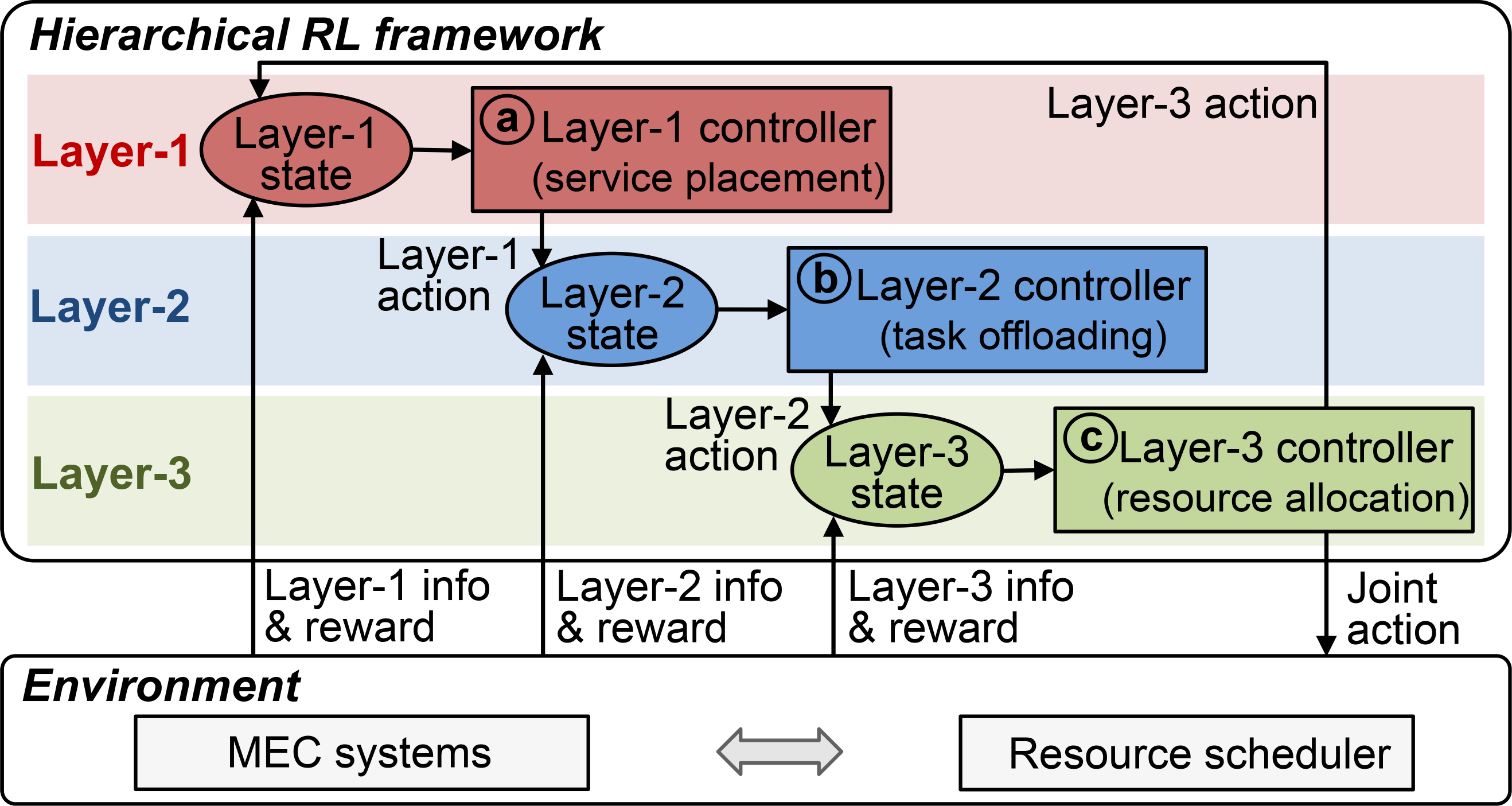}}
\caption{The workflow of three-layer hierarchical DRL framework.}
\vspace{-1em}
\label{hrl overview}
\end{figure}
Hierarchical deep reinforcement learning (HDRL) \cite{kulkarni2016hierarchical} is a promising framework to extend traditional DRL algorithms to learn complicated tasks, by breaking the overall decision-making down into a hierarchy of sub-tasks.
As shown in Fig. \ref{hrl overview}, we design a tailored HDRL framework that comprises three-layer DRL controllers, i.e., \textcircled{a} Layer-1 controller, \textcircled{b} Layer-2 controller, and \textcircled{c} Layer-3 controller, to cope with three layers of scheduling sub-problems $\mathcal{P}$1, $\mathcal{P}$2, and $\mathcal{P}$3, respectively.

At slot $t$, the input of each layer of the controllers is the layer-specific state composed of two components, i.e., the status of queueing tasks after executing the action of the layer above it and the current resource configuration information from the environment.
The DRL controller uses the state to select the action that determines whether to update the current scheduling decisions of the corresponding layer.

After performing the joint three-layer action, the environment transitions to slot $t+1$ with new three-layer scheduling information, and provides three-layer controllers with respective rewards.
The reward for each controller is set based on the targeted sub-problem's objective.
The controllers use these rewards to improve their actions for future time slots.
Details on states, actions, and rewards are provided in subsection \ref{MARL}.

\subsection{Safe Multi-Agent DRL for Decentralized Scheduling}
\label{MARL}
Each layer of the controllers uses the tailored multi-agent DRL algorithm to generate decentralized actions, and a safe learning scheme is applied to ensure system reliability.

\emph{1) Decentralized POMDP Modeling:}
\begin{table*}[!t]
\caption{Settings of DRL components for each layer of the controllers considered.}
\vspace{-1em}
\begin{center}
\begin{tabular}{m{93pt}m{135pt}m{75pt}m{158pt}}
\hline
\textbf{Layers of DRL controllers} & \textbf{States of each agent} & \textbf{Actions of each agent} & \textbf{Rewards for all agents}\\
\hline
\multicolumn{1}{c}{\makecell[l]{Layer-1 controller\\edge-cloud service placement}}& extent of service coverages, available memory, installed service types, service types of arrived tasks & whether to update the current placement decisions of its own agent& $\makecell[l]{\sum_{i\in\mathcal{N}}[(a\cdot \text{PlaceCost} _{i}\\+b\cdot(\text{TaskUnserved} _{i}-\text{DudgetUnserved}))^{-1}]}$\\
\hline
\multicolumn{1}{c}{\makecell[l]{Layer-2 controller\\edge-edge task offloading}}& ratio of tasks that violate delay budgets, available CPU unit, amount of unoffloaded workloads & whether to update the current offloading decisions of its own agent& $\makecell[l]{\sum_{i\in\mathcal{N}}[(c\cdot\text{OffloadCost} _{i}+d\cdot (\text{TaskDelay} _{i}\\-\text{DudgetDelay}))^{-1}]+e\cdot \text{WorkloadVariance}}$\\
\hline
\multicolumn{1}{c}{\makecell[l]{Layer-3 controller\\intra-edge resource allocation}}& ratio of tasks that violate delay budgets, rate variance among tasks, available CPU unit, workloads waiting for resource allocation& whether to update the current allocation decisions of its own agent& $\makecell[l]{\sum_{i\in\mathcal{N}}[\text{UndelayedWorkload}_{i}/(f\cdot \text{AllocateCost} _{i}\\+g\cdot (\text{TaskDelay} _{i}-\text{DudgetDelay}))]}$\\
\hline
\end{tabular}
\label{settings}
\end{center}
\end{table*}
To save the communication cost, each of the controllers is required to perform actions in a decentralized manner, where \emph{each edge server makes its own decisions independently without relying on the global system information from other servers.}
We formulate the learning process as a decentralized partially observable Markov decision process (Dec-POMDP) \cite{oliehoek2012decentralized}.
For each of the controllers, \emph{each edge server acts as an agent} to learn decentralized actions.
In addition to conventional components of a MDP, such as a state space $\mathcal{S}$, an action space $\mathcal{A}$, a state transition function $p: \mathcal{S}\times\mathcal{A}\rightarrow\mathcal{S}$, a shared reward function $r: \mathcal{S}\times\mathcal{A}\rightarrow\mathbb{R}$, and a discount factor $\gamma$, an extra local observation space $\mathcal{O}= \times_{i\in\mathcal{N}}\mathcal{O}_{i}$ is considered in Dec-POMDP to reflect local observations available to each agent from the global state $\mathcal{S}$. 

Each agent $i$ uses its policy $\pi_{i}(a_{i}(t)|o_i(t))$ to select the action $a_{i}(t)$ from the partial observations $o_i(t)$.
The goal of the Dec-POMDP is to find a joint policy of all agents $\Pi=\{\pi_{i}(a_{i}(t)|o_i(t))|i\in\mathcal{N}\}$ that maximizes the expected cumulative reward:
\begin{align}
\setlength{\abovedisplayskip}{-1pt}
\setlength{\belowdisplayskip}{-1pt}
&\max_{\Pi} \mathbb E[\sum_{t=0}^{\infty} \gamma^t r(\mathcal{S}(t), \Pi(\mathcal{A}(t)|\mathcal{O}(t)))],
\label{dec-pomdp}
\end{align}
where $\mathcal{S}(t)$, $\mathcal{A}(t)$, and $\mathcal{O}(t)$ are global states, actions, and observations of all agents at slot $t$, respectively.
The layer-specific state, action, and reward are summarized in Table \ref{settings}.

\emph{2) Safe DRL Design:}
In our problem, DRL may generate unsafe actions that are in conflict with the executed actions. 
For example, the Layer-2 action may result in tasks not being offloaded due to the fact that the Layer-1 controller has just removed the required services from the destination server.
The unsafe actions result in high scheduling delay for these unoffloaded tasks.
To ensure safety in DRL, 
we design a \emph{discriminator-aided action masking} scheme to avoid executing unsafe actions.
Specifically, we build a discriminator to check whether the current action meets safety conditions, e.g., the destination server has installed the corresponding services.
For unsafe actions, their associated logits, i.e, raw outputs of neural networks, are replaced by large negative numbers, thereby the probability of selecting these unsafe actions becomes virtually zero.
In this way, agents will replace the original actions with ”update decision” actions before applying them to the system.
Then, EdgeTimer will use built-in scheduling rules, e.g., Kubernetes rules \cite{k8s}, to determine safe scheduling results within the feasible region.

\emph{3) Algorithm Training for Safe Multi-Agent DRL:}
\begin{figure}[!t]
\centerline{\includegraphics[width=0.95\columnwidth]{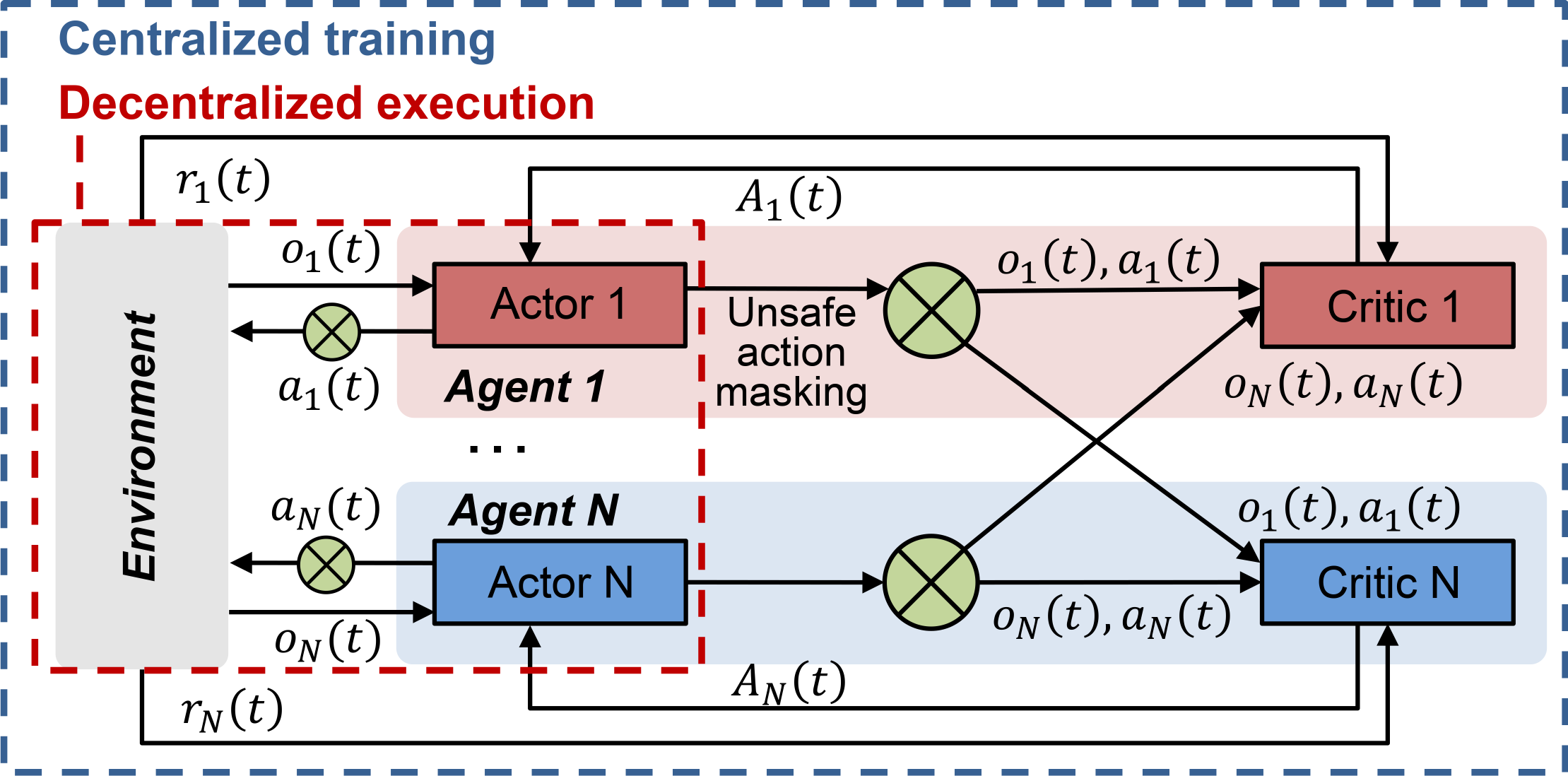}}
\caption{The illustration of safe multi-agent DRL algorithm.}
\vspace{-1em}
\label{MARL overview}
\end{figure}
\begin{algorithm}[!t]
\small
\caption{EdgeTimer for adaptive multi-timescale resource scheduling}
\label{training algorithm}
\textbf{$/*$ Initialization $*/$}\\
\lnl{InRes2} Initialize parameters of neural networks, actions ($a_{i}^{1}(0), \!a_{i}^{2}(0), \!a_{i}^{3}(0)$), and replay buffers ($R^{1}, \!R^{2}, \!R^{3}$);\\
\lnl{InRes2} Initialize the network configuration $\widetilde{s}$;\\
\lnl{InRes4} \For{\rm{epoch} $=0, \cdots, M$}{
\lnl{InRes4} \For{\rm{time slot} $t=1, \cdots, T$}{
\lnl{InRes7} \For{\rm{each edge server} $i =1, \cdots, N$}{
\textbf{$/*$ Generate Layer-1 actions $*/$}\\
\lnl{InRes7} $o^{1}_{i}(t) \gets$ \Call{Step}{$a^1_{i}(t-1), a^3_{i}(t-1)$};\\
\lnl{InRes7} $a^{1}_{i}(t) \gets$ $\pi(o^{1}_{i}(t)|\theta_{i}^{1})$;\\
\textbf{$/*$ Generate Layer-2 actions $*/$}\\
\lnl{InRes7} $o^{2}_{i}(t) \gets$ \Call{Step}{$a^2_{i}(t-1), a^1_{i}(t)$};\\
\lnl{InRes7} $a^{2}_{i}(t) \gets$ $\pi(o^{2}_{i}(t)|\theta_{i}^{2})$;\\
\textbf{$/*$ Generate Layer-3 actions $*/$}\\
\lnl{InRes7} $o^{3}_{i}(t) \gets$ \Call{Step}{$a^3_{i}(t-1), a^2_{i}(t)$};\\
\lnl{InRes7} $a^{3}_{i}(t) \gets$ $\pi(o^{3}_{i}(t)|\theta_{i}^{3})$;\\
}
\textbf{$/*$ Compute three-layer rewards $*/$}\\
\lnl{InRes7} Perceive rewards $r^{1}_{i\in\mathcal{N}}(t)$, $r^{2}_{i\in\mathcal{N}}(t)$, and $r^{3}_{i\in\mathcal{N}}(t)$;\\
\lnl{InRes7} Store transitions of three controllers in $R^{1}$, $R^{2}$, and $R^{3}$, respectively;\\
\lnl{InRes7} \If{\rm {\Call{Done}{$o^{1}_{i\in\mathcal{N}}(t), o^{2}_{i\in\mathcal{N}}(t), o^{3}_{i\in\mathcal{N}}(t)$}}}{
\lnl{InRes7} $o^{1}_{i\in\mathcal{N}}(t), o^{2}_{i\in\mathcal{N}}(t), o^{3}_{i\in\mathcal{N}}(t) \gets$ \Call{Reset}{$\widetilde{s}$};\\
}
}
\lnl{InRes7}Sample mini-batch $W^{1}$, $W^{2}$, and $W^{3}$ from $R^{1}$, $R^{2}$, and $R^{3}$ as training samples, respectively;\\
\lnl{InRes7}Update parameters of networks using (\ref{actor}) and (\ref{critic});
}
\end{algorithm}
We integrate safe learning with multi-agent DRL (MADRL) \cite{yu2022surprising} to achieve the goal of Dec-POMDP in (\ref{dec-pomdp}).
As shown in Fig. \ref{MARL overview}, we adopt a \emph{centralized training and decentralized execution} scheme, allowing agents to use non-local information from other agents to ease offline training, while \emph{the extra information is not allowed to use during the online inference}.
Specifically, each agent has two neural networks, where an actor network learns the policy with local observations, and a critic network approximates the centralized value function from the global environment to evaluate the current policy.
The training process is shown in Algorithm \ref{training algorithm}.

\textbf{Decentralized execution with actors.}
For each agent $i\in{\mathcal{N}}$, its actor $i$ maps a local observation $o_{i}(t)$ to an action $a_{i}(t)$ with policy $\pi_{\theta_i}$ parameterized by $\theta_{i}$.
To maximize the return obtained by the policy in (\ref{dec-pomdp}), the actor for agent $i$ is updated following the policy gradient theorem \cite{sutton1999policy}:
\begin{equation}
\nabla_{\theta_i}\!J(\!\pi_{\theta_i}\!)\!=\!\mathbb{E}_{\pi_{\theta_i}}\![\nabla_{\theta_i}\!\log {\pi_{\theta_i}}\!(a_i(t)|o_i(t))\!A_i(\!\mathcal{O}(t),\!\mathcal{A}(t))],
\label{actor}
\end{equation}
where $A_i(\mathcal{O}(t),\mathcal{A}(t))$ is the centralized advantage function that receives global observations, $\mathcal{O}(t)=(o_1(t),...,o_N(t))$, together with actions of all agents, $\mathcal{A}(t)=(a_1(t),..,a_N(t))$, and outputs a measure of policy performance for agent $i$.
The advantage function can be computed using generalized advantage estimation (GAE) \cite{schulman2015high}:
\begin{equation}
\setlength{\belowdisplayskip}{-1pt}
A_i(\mathcal{O}(t),\mathcal{A}(t))=\sum_{l=0}^{T-t}(\gamma\lambda)^{l}\overline{V}_{i}(t+l),
\end{equation}
\begin{equation}
\setlength{\abovedisplayskip}{-1pt}
\setlength{\belowdisplayskip}{-1pt}
\overline{V}_{i}(t+l)= r_{i}(t+l)+\gamma V_{i}(\mathcal{O}(t+l+1))-V_{i}(\mathcal{O}(t+l)),
\end{equation}
where $\lambda$ is the weight discount to control the trade-off between variance and bias, $T$ is the trajectory length, and $V_i(\mathcal{O}(t))$ is the value function that indicates the expected return of starting from global observations $\mathcal{O}(t)$ under policy $\pi_{\theta_i}$.
We design an individual critic network for agent $i$ to approximate $V_i(\mathcal{O}(t))$.

\textbf{Centralized training with critics.}
The critic $V_{\omega_i}$ for agent $i$ with parameter $\omega_i$ can be trained by minimizing the mean-square error as follows:
\begin{equation}
\setlength{\abovedisplayskip}{-1pt}
\setlength{\belowdisplayskip}{-1pt}
L(\omega_i)=(V_{\omega_i}(\mathcal{O}(t))-\sum_{k=0}^{T-t}\gamma^{k}r_{i}(t+k))^{2}.
\label{critic}
\end{equation}
\section{Implementation}
\label{implementation}
EdgeTimer was implemented on a realistic Vienna system-level simulator \cite{muller2018flexible}, embedding $45$ combinations of typical scheduling rules for three-layer scheduling.
\subsection{MEC System Simulator}
We implemented $12$ edge servers and one cloud server in a realistic Vienna system-level simulator\footnote{The simulator enables to perform resource scheduling of 5G multi-tier networks. We extended it to MEC system scenarios by considering each base station as an edge server and adding an extra network node as a cloud.} \cite{muller2018flexible} with MATLAB.
In the simulation, edge servers were distributed over a $5$ km$\times$$5$ km region.
Each edge server was equipped with four CPU cores and eight GB memory.
The cloud had eight CPU cores and $16$ GB memory.
The length of each time slot was one second.
The unit price of computation resource was derived from the pay-as-you-go pricing model at \!Microsoft \!Azure \cite{microsoft}.
To simulate operation costs, we measured the service migration time \cite{guo2023efficient}, handover time \cite{sun2017emm}, and server lead time \cite{jiao2017smoothed} in service placement, task offloading, and resource allocation, respectively, and converted them to the cost based on the hourly price at \!Microsoft \!Azure. 
For simplicity, we dropped the dependency on the money and service types and used relative weights to set the unit price, placement cost, offloading cost, and unit allocation cost as $25$, $0.3d_1$, $0.1d_2$, and $0.5$, respectively. \!Parameters \!$d_1$ and \!$d_2$ are the transmission distance of service migration and task offloading, respectively.
\subsection{Integration With Representative Scheduling Rules}
\label{built-in}
The simulator embedded a total of $45$ combinations of representative scheduling rules for service placement, task offloading, and resource allocation.
These built-in rules were executed when EdgeTimer generated the "update decision" action (discussed in subsection \ref{problem formulation sub}). 
The service placement rules included a Kubernetes placement rule-horizontal pod autoscaling (HPA) \cite{k8shpa}, Top-K rule \cite{farhadi2019service}, and always migration rule (AM) \cite{ma2020leveraging}.
For task offloading, we implemented 5 offloading rules of Kubernetes \cite{k8s} (MRP, LRP, RLP, SSP, and RCRP).
The allocation rules consisted of proportional fairness (PF) \cite{kelly1997charging}, round robin (RR) \cite{aziz2019constrained}, and equal allocation (EA).
\subsection{Scalability to Multi-Pattern Workloads}
When applying DRL to task requests with varying patterns, sparse reward issues may arise, i.e., the agents rarely receive useful reward signals.
For instance, when an agent selected actions during the idle period with empty queues, e.g., the night period, it would receive invalid rewards.
Instead, we designed an automatic skip mechanism that allowed agents to use non-DRL-aided default actions when the request queue was empty.
The default action of each edge server was set to continue its previous scheduling decisions.
\subsection{Training Settings}
The safe multi-agent DRL algorithm was implemented in Python 3.6.2 with PyTorch 1.10.2.
Both the actor and critic networks included a one-layer RNN with 64 hidden units and were updated by the learning rate of $5\times10^{-4}$.
The discount factor $\gamma$ was set to $0.99$, and parameter $\lambda$ in GAE was $0.95$. 
\section{Evaluation}
\label{evaluation}

\subsection{Methodology}
The training and inference were conducted with an Intel(R) Core CPU (2.90GHz 8-cores) and 512GB RAM.
\begin{figure}[t!]
    \begin{minipage}{0.49\linewidth} 
      \includegraphics[width=1\columnwidth]{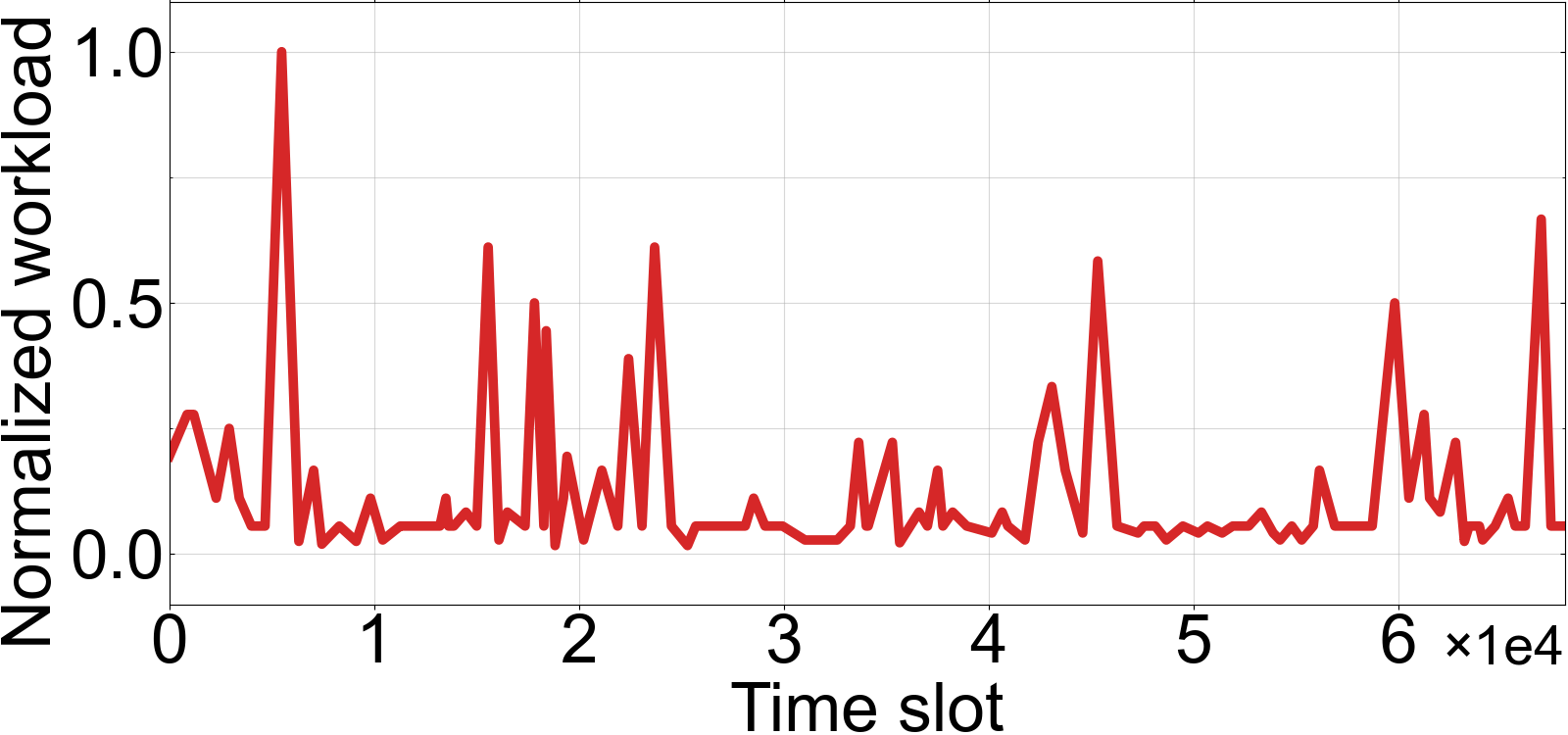} \\\footnotesize{(a) Pattern A: raw workload.}
    \end{minipage}
    \begin{minipage}{0.49\linewidth} 
      \includegraphics[width=1\columnwidth]{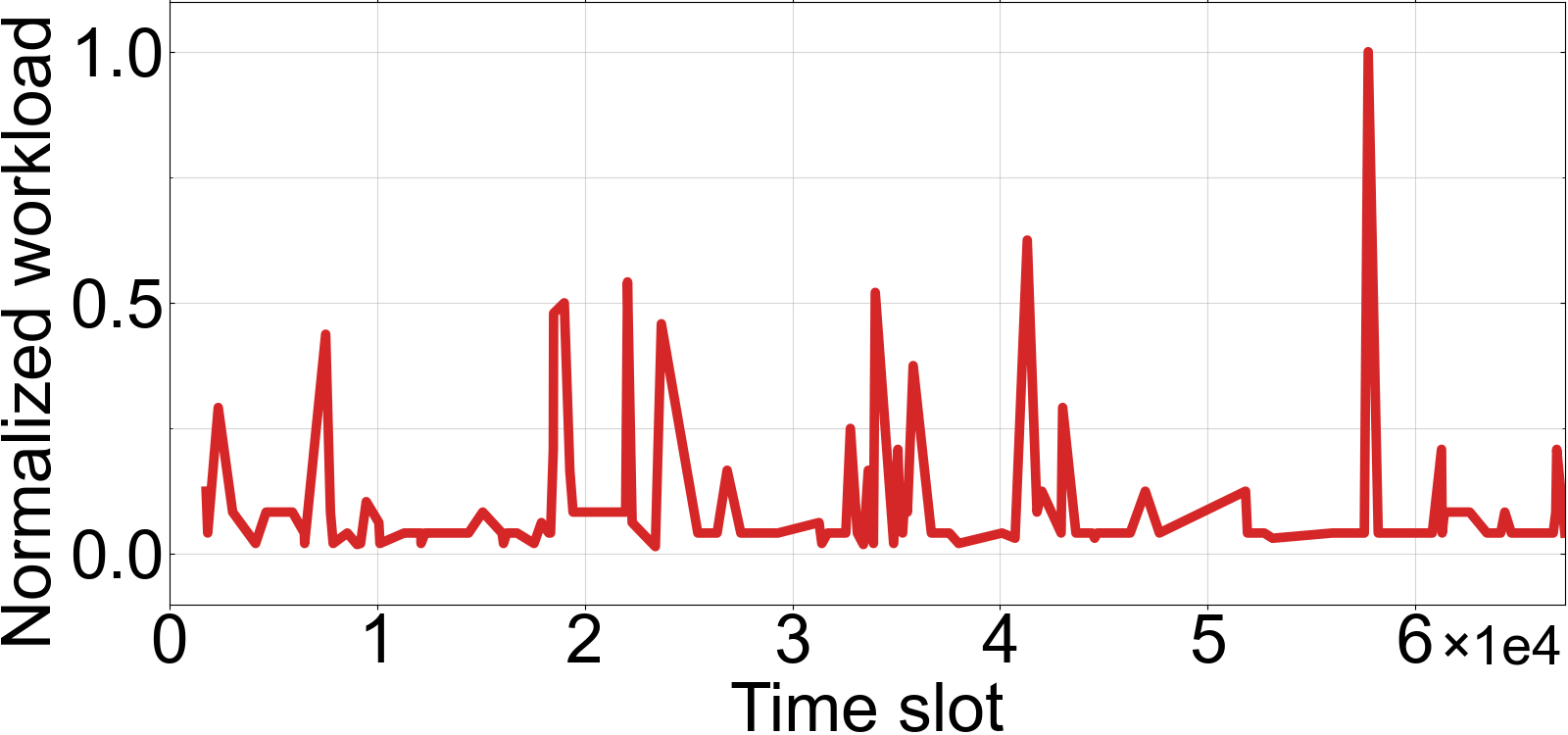} \\\footnotesize{(b) Pattern B: request shuffle.}
    \end{minipage}
    \begin{minipage}{0.49\linewidth} 
      \includegraphics[width=1\columnwidth]{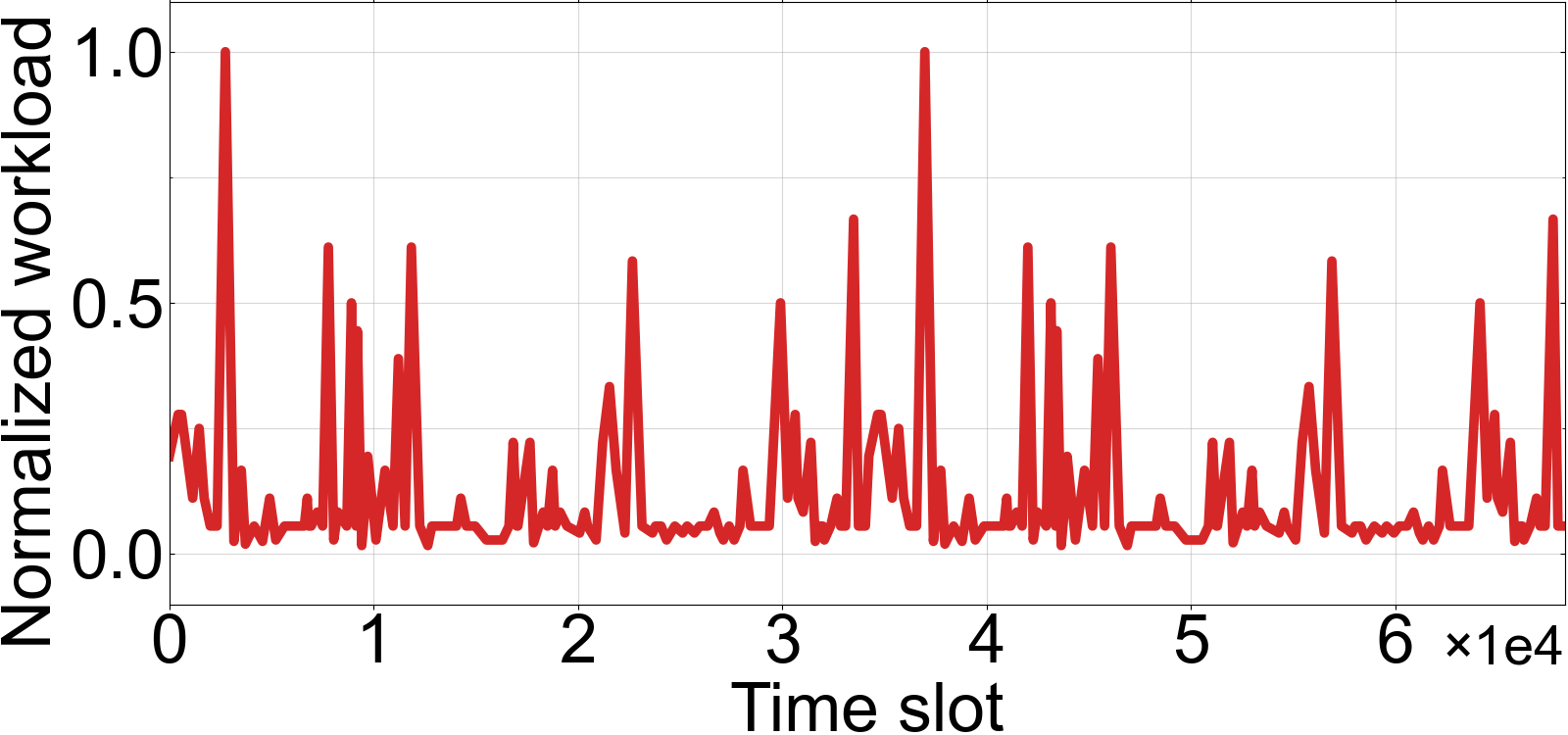} \\\footnotesize{(c) Pattern C: $2\times$ request frequency.}
    \end{minipage}
    \begin{minipage}{0.49\linewidth} 
      \includegraphics[width=1\columnwidth]{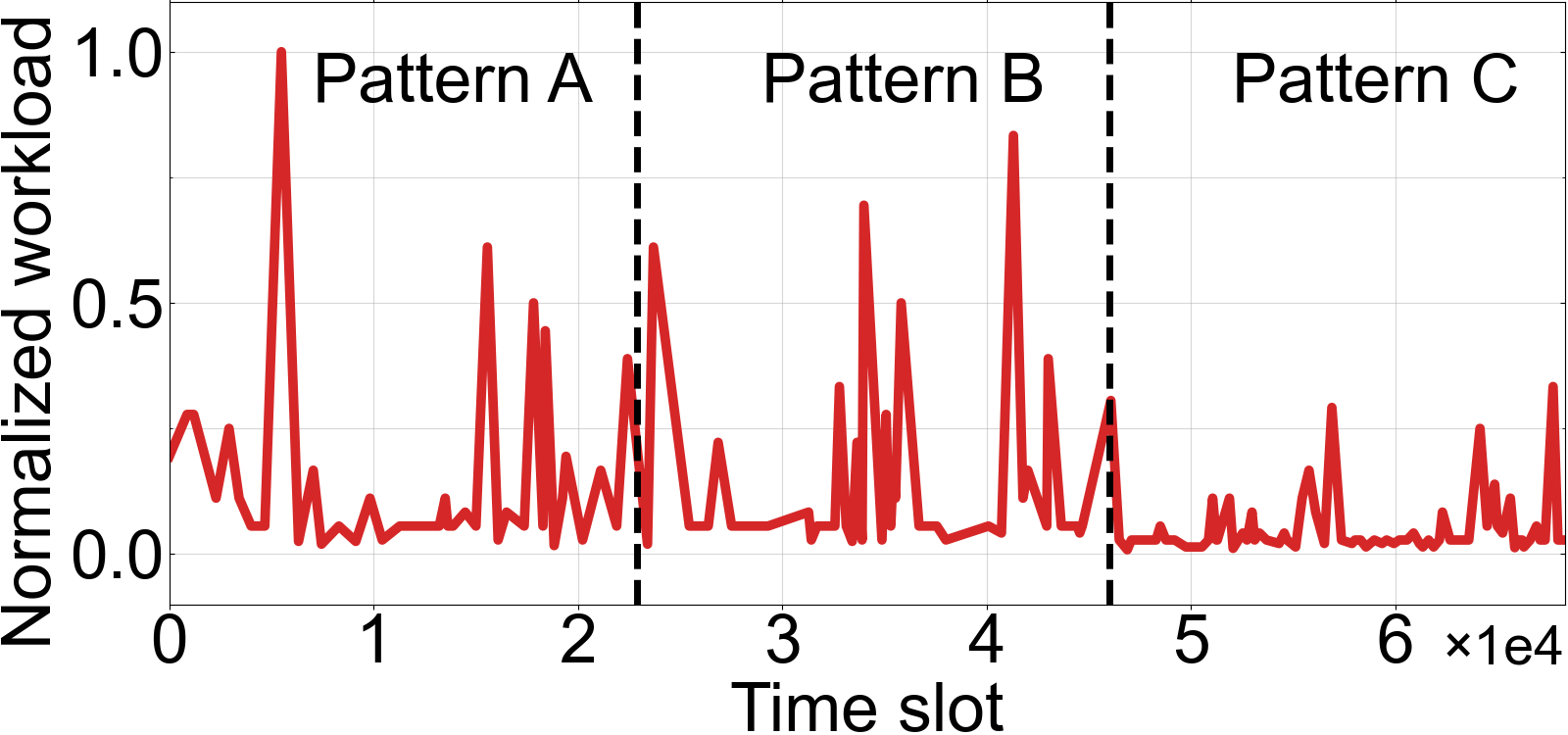} \\\footnotesize{(d) Pattern D: A-B-C combination.}
    \end{minipage}
  \caption{Illustrations of four workload patterns.}
  \vspace{-1em}
  \label{multi-pattern trace}
\end{figure}

\textbf{Multi-Pattern Task Workloads.}
We evaluated EdgeTimer on real-world cluster traces from Alibaba \cite{Ali} that provided task information per second within 8 days, including start time, end time, task type, required CPU, and required memory.
We considered "task type" in traces as the type of services, and divided task requests into $12$ services.
The workload of each service task was estimated by multiplying the number of CPU cores it required by its actual processing time.
Following the methodology in KaiS \cite{han2021tailored}, the delay budget of each service task was set by proportionally scaling its actual processing time.
To evaluate EdgeTimer in task arrivals with different patterns, we generated multi-pattern trace as shown in Fig. \ref{multi-pattern trace}.
The pattern A was the raw workload under stochastic arrivals (Fig. \ref{multi-pattern trace}(a)).
In pattern B, we shuffled the order in which the edge server received tasks to reflect interference from other tasks (Fig. \ref{multi-pattern trace}(b)).
Furthermore, we doubled the request frequency in pattern C to simulate an increase in workloads (Fig. \ref{multi-pattern trace}(c)), and concatenated the above three patterns as pattern D to evaluate the adaptability to pattern changes (Fig. \ref{multi-pattern trace}(d)).

\textbf{Comparatives.}
We compared EdgeTimer with four comparison algorithms: \emph{Static single-timescale (SST)} \cite{poularakis2019joint, ma2020cooperative} that updates three-layer scheduling decisions at each slot; \emph{Static multi-timescale (SMT)} \cite{farhadi2019service, han2021tailored} that updates decisions of each layer via a fixed and specific timescale\footnote{For each layer, we iterated through the timescale set, $\{1,10,50,100\}$, to select the optimal as its fixed timescale.}; \emph{Delay-triggered update (DT)} and \emph{workload-triggered update (WT)} that update an edge's decisions when the delay and waiting workload in the edge exceed corresponding thresholds, respectively. 
\subsection{Performance on Multi-Pattern Workloads}
\label{Performance on Multi-Pattern Workloads}
\begin{figure}[t!]
    \begin{minipage}{0.49\linewidth} 
      \includegraphics[width=1\columnwidth]{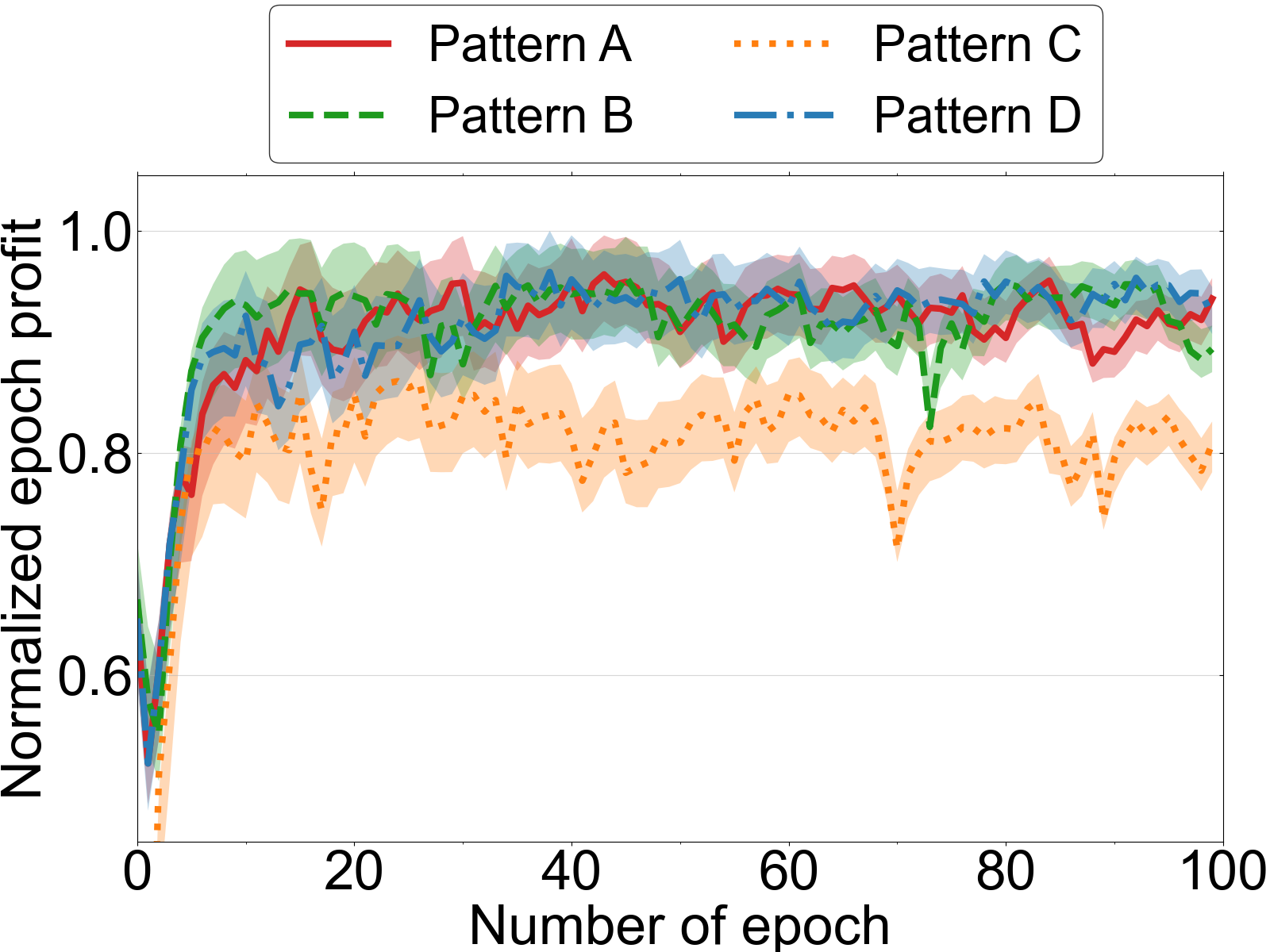} \\\centerline{\footnotesize{(a)}}
    \end{minipage}
    \begin{minipage}{0.49\linewidth} 
      \includegraphics[width=1\columnwidth]{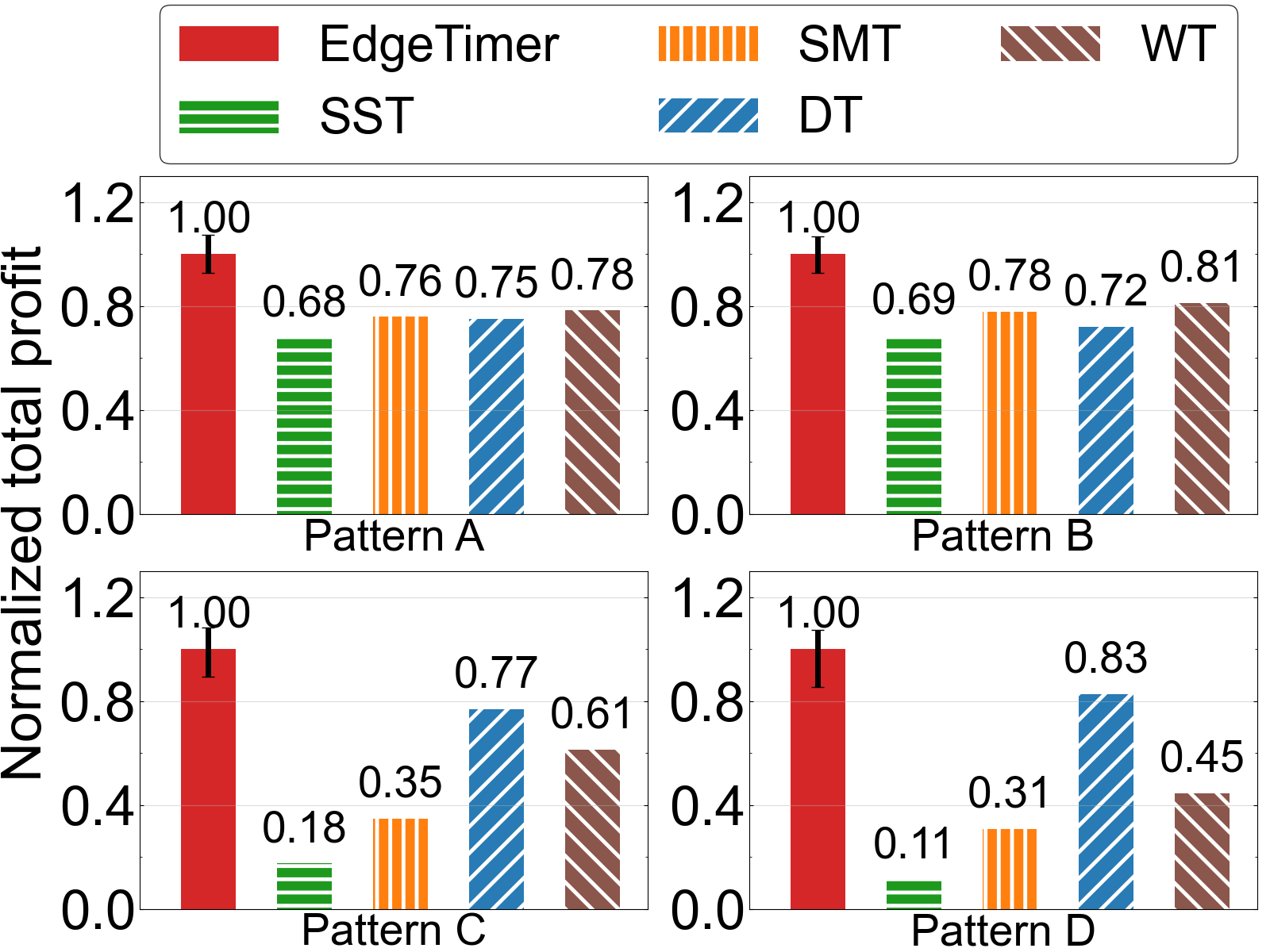} \\\centerline{\footnotesize{(b)}}
    \end{minipage}
  \caption{(a) Normalized epoch profit of EdgeTimer and (b) normalized total profit of EdgeTimer and comparatives in different workload patterns.}
  \label{profit with baselines}
\end{figure}
We compared EdgeTimer with comparatives in four workload patterns (Fig. \ref{multi-pattern trace}).
The built-in scheduling rule used in this subsection is AM-MRP-EA, i.e., AM for service placement, MRP for task offloading, and EA for resource allocation.

\textbf{Profit Performance.}
Fig. \ref{profit with baselines}(a) shows the normalized epoch profit of EdgeTimer versus the number of offline training epochs in different workload patterns.
The epoch profit is normalized against the profit obtained in pattern D.
It shows that EdgeTimer enables to support different workload patterns and converges within $20$ epochs.
Even for pattern C and D where the learning task becomes more intractable due to the workload spike and the pattern fluctuating, nearly equivalent epochs are required for convergence compared with that in the raw trace.
The converged profit is lowest in pattern C.
This is because as the request frequency increases, more operation costs are required to complete tasks within the delay budget.

Fig. \ref{profit with baselines}(b) demonstrates the online profit performance for EdgeTimer and four comparatives.
The total profit is normalized against that of EdgeTimer in each pattern.
EdgeTimer achieves the highest profit, irrespective of workload patterns.
For pattern A, EdgeTimer stably improves the profit for $1.47\times$, $1.32\times$, $1.33\times$, and $1.28\times$ than SST, SMT, DT, and WT approaches, respectively.
In particular, EdgeTimer shines in high-workload patterns (pattern C) and high-fluctuating (pattern D), where EdgeTimer achieves $5.56\times$ and $9.1\times$ higher profit than comparatives, respectively.

\begin{figure}[!t]
\centerline{\includegraphics[width=1\columnwidth]{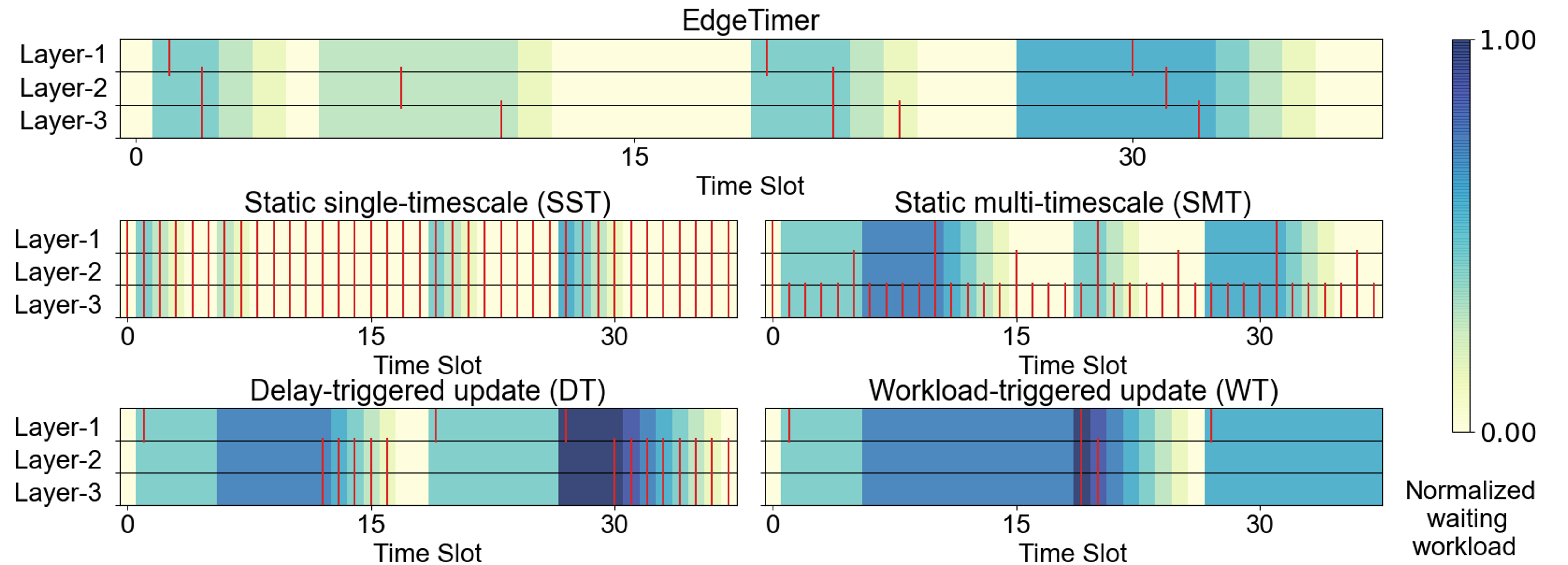}}
\caption{Illustrations of updating timescales versus normalized waiting workloads for EdgeTimer and four comparatives. The scheduling decisions are updated at red vertical lines.}
\vspace{-1em}
\label{the reason for the performance}
\end{figure}
Fig. \ref{the reason for the performance} investigates the underlying efforts behind EdgeTimer.
It shows that the effectiveness of EdgeTimer comes from adaptively adjusting the update frequency of scheduling decisions based on the current task performance, compared with SST and SMT.
Furthermore, EdgeTimer outperforms DT and WT, as it enables to mitigate potential overload in advance using more comprehensive observations, rather than relying solely on the delay or workload metrics (dense red vertical lines only appear at the end or beginning of the dark purple blocks).

\textbf{Delay Performance.}
\begin{figure}[t!]
    \begin{minipage}{0.27\linewidth} 
      \includegraphics[width=1\columnwidth]{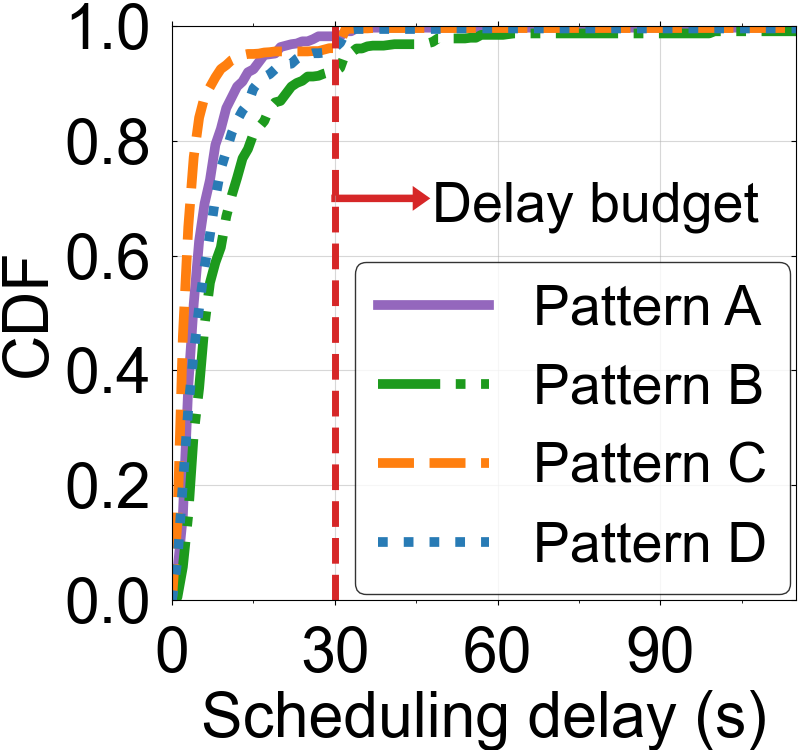} \\\centerline{\footnotesize{(a)}}
    \end{minipage}
    \begin{minipage}{0.39\linewidth} 
      \includegraphics[width=1\columnwidth]{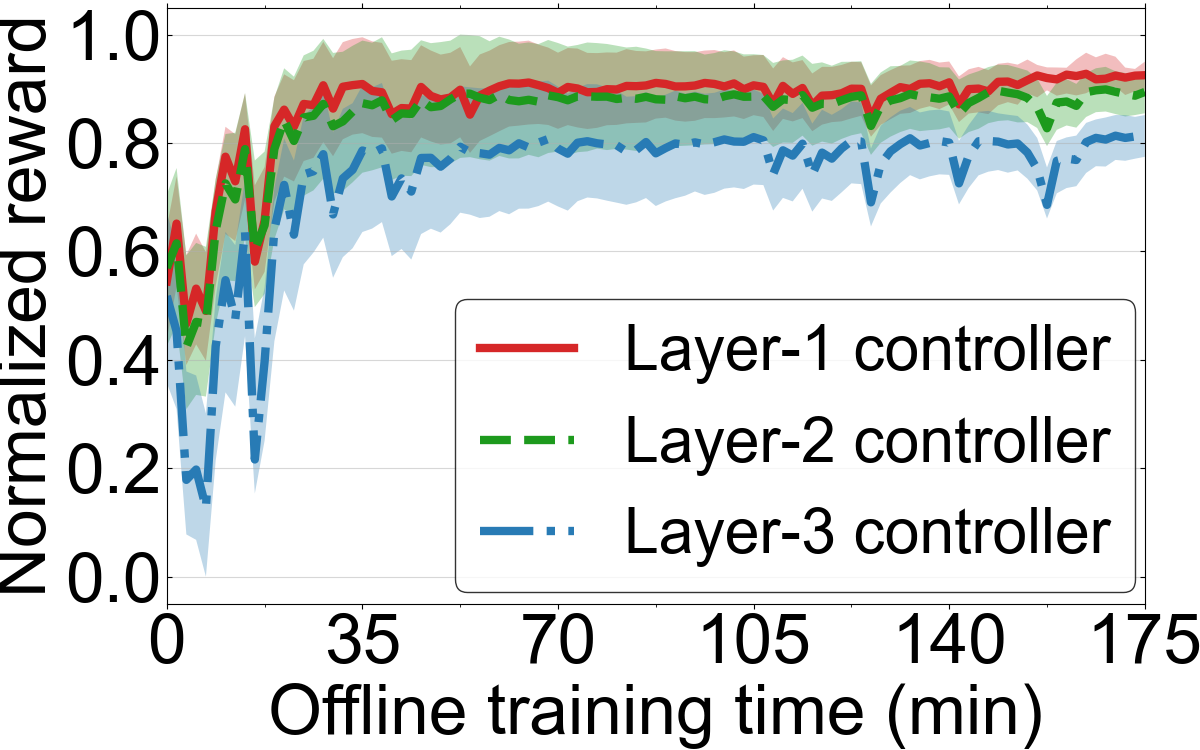} \\\centerline{\footnotesize{(b)}}
    \end{minipage}
    \begin{minipage}{0.27\linewidth} 
      \includegraphics[width=1\columnwidth]{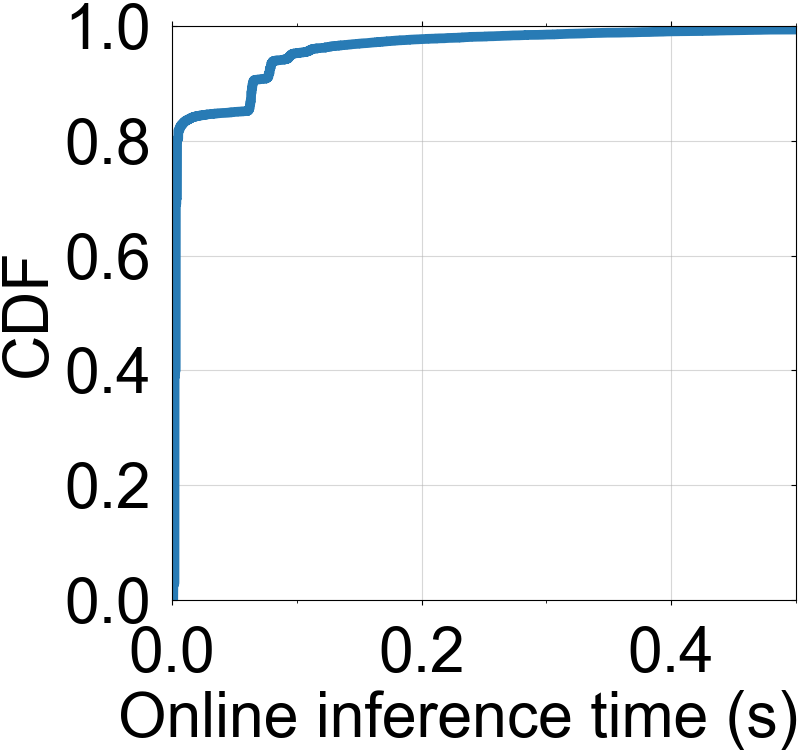} \\\centerline{\footnotesize{(c)}}
    \end{minipage}
  \caption{(a) CDF of the scheduling delay in EdgeTimer under different patterns. (b) Normalized rewards of each layer of DRL controllers in EdgeTimer. (c) CDF of the online inference time in EdgeTimer.}
  \label{CDF of the time}
\end{figure}
Given a service type, Fig. \ref{CDF of the time}(a) shows the CDF of the scheduling delay for such service requests under different workload patterns.
We can observe that nearly all service requests can be completed within the delay budget, irrespective of workload patterns.
It indicates that huge profits in EdgeTimer are not achieved at the expense of the user experience, i.e., delay performance.     

\textbf{Training and Inference Time.}
Fig. \ref{CDF of the time}(b) demonstrates offline average rewards of Layer-1, Layer-2, and Layer-3 controllers in EdgeTimer.
The overall controllers enable to achieve convergence within $35$ minutes.
Fig. \ref{CDF of the time}(c) shows the CDF of the time spent by EdgeTimer to generate policies during the online process.
The average inference time is $0.026$ second that is less than the typical time interval between scheduling events, i.e, one second.
These results indicate that EdgeTimer manages to execute scheduling policies at each time slot and imposes no extra overhead in production. 

\subsection{Performance With Different Built-In Scheduling Rules}
\label{Built-In Scheduling Rules}
\begin{table*}[t!]
 \renewcommand{\arraystretch}{1.2}
 \setlength\tabcolsep{3pt} 
 \caption{Profit gains compared to existing approaches under different built-in scheduling rules in pattern A}
 \vspace{-1em}
 \begin{center}
  \begin{tabular}{|c|c|c|c|c|c|c|c|c|c|c|}
   \hline
   \multicolumn{2}{|c|}{\multirow{2}{*}{}}  & \multicolumn{9}{c|}{Service placement rule - resource allocation rule} \\
   \cline{3-11}
   \multicolumn{2}{|c|}{~} & HPA - PF & HPA - RR & HPA - EA & Top-K - PF & Top-K - RR & Top-K - EA & AM - PF & AM - RR & AM - EA \\
   \hline
   \multirow{5}{0.15\columnwidth}{\centering Task offloading rule} & MRP & 1.73-2.43x & 1.70-2.39x & 1.66-2.81x & 1.40-3.87x & 1.39-3.86x & 1.40-3.55x & 1.30-1.55x & 1.31-1.56x & 1.28-1.47x \\
   \cline{2-11}
   & LRP & 1.71-2.39x & 1.70-2.39x & 1.72-2.94x & 1.34-3.85x & 1.36-3.90x & 1.53-3.56x & 1.26-1.53x & 1.25-1.52x & 1.26-1.48x \\
   \cline{2-11}
   & RLP & 1.72-2.41x & 1.71-2.40x & 1.72-2.92x & 1.31-3.83x & 1.32-3.87x & 1.48-3.60x & 1.26-2.37x & 1.27-2.38x & 1.33-1.49x \\
   \cline{2-11}
   & SSP & 1.38-1.99x & 1.39-2.01x & 1.41-2.98x & 1.91-4.67x & \textbf{1.92-4.70x} & 1.29-3.61x & 1.28-3.89x & 1.26-3.83x & 1.23-2.07x \\
   \cline{2-11}
   & RCRP & 1.72-2.41x & 1.73-2.43x & 1.70-2.92x & 1.32-3.82x & 1.30-3.78x & 1.57-3.63x & 1.26-1.59x & 1.26-1.60x & 1.30-1.50x \\
   \hline
  \end{tabular}
  \label{built-in rules}
 \end{center}
\end{table*}
Table \ref{built-in rules} shows profit gains compared with existing approaches under $45$ built-in scheduling rules in pattern A.
EdgeTimer outperforms existing approaches for all built-in scheduling rules.
For some Kubernetes rules, such as HPA-MRP-PF, EdgeTimer improves the profit performance for up to $2.43\times$ compared with approaches under the same rule. 
\subsection{EdgeTimer Deep Dive}
\label{EdgeTimer Deep Dive}
\textbf{Impact of Three-Layer Hierarchical DRL Design.}
\begin{figure}[t!]
    \begin{minipage}{0.49\linewidth} 
      \includegraphics[width=1\columnwidth]{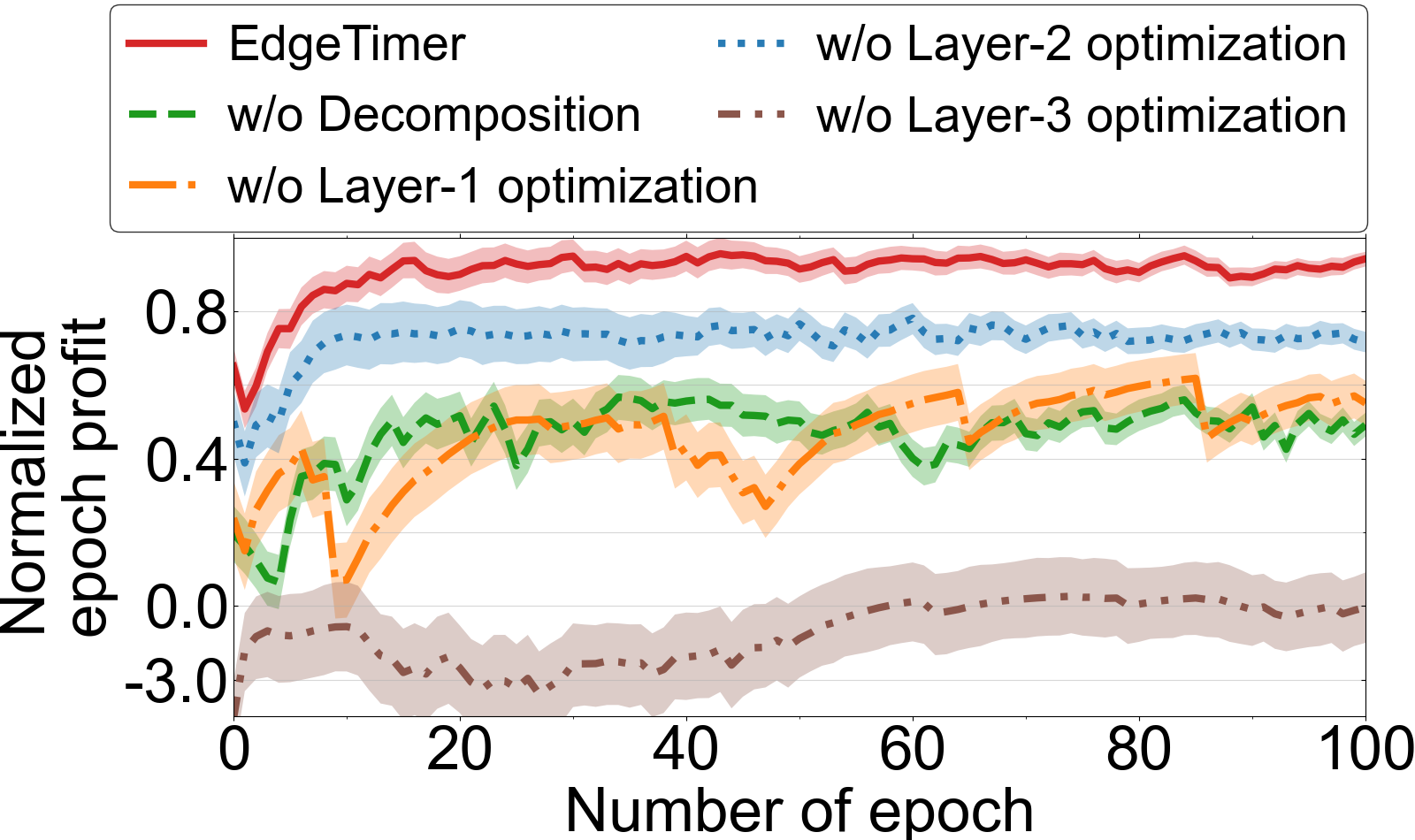} \\\footnotesize{(a) Offline profits.}
    \end{minipage}
    \begin{minipage}{0.49\linewidth} 
      \includegraphics[width=1\columnwidth]{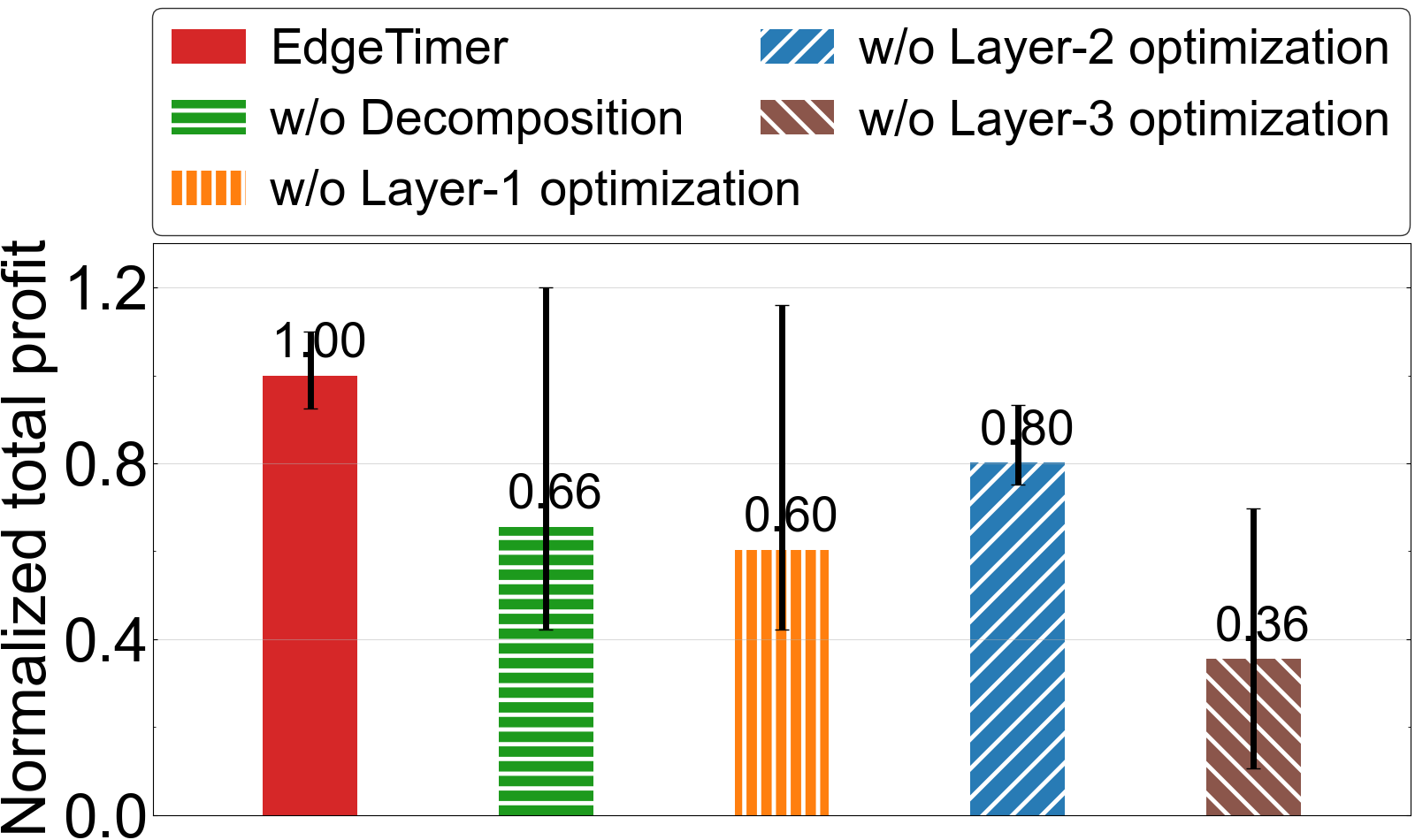} \\\footnotesize{(b) Online profits.}
    \end{minipage}
  \caption{Impact of the three-layer hierarchical DRL design on profit results.}
  \vspace{-1em}
  \label{Impact of Hierarchical}
\end{figure}
We design a three-layer hierarchical DRL framework to decompose the original task into three layers of sub-tasks and optimize them in a coordinated manner.
To verify its contribution, Fig. \ref{Impact of Hierarchical} compared the offline and online profit of EdgeTimer, EdgeTimer without task decomposition, i.e, w/o Decomposition, and EdgeTimer without each layer of sub-task optimization in turn, i.e, w/o Layer-1 optimization, w/o Layer-2 optimization, and w/o Layer-3 optimization.
All results are normalized against that of EdgeTimer.
EdgeTimer accelerates the convergence in the offline process and obtains $1.52\times$ higher online profit than w/o Decomposition, since the task decomposition scheme simplifies the learning task significantly.
Moreover, removing any layer of optimization from EdgeTimer results in a decrease in profit of at least $1.25\times$, which reveals that each layer is indispensable for the overall performance.

\textbf{Impact of Safe Multi-Agent DRL Design.}
\begin{figure}[t!]
    \begin{minipage}{0.49\linewidth} 
      \includegraphics[width=1\columnwidth]{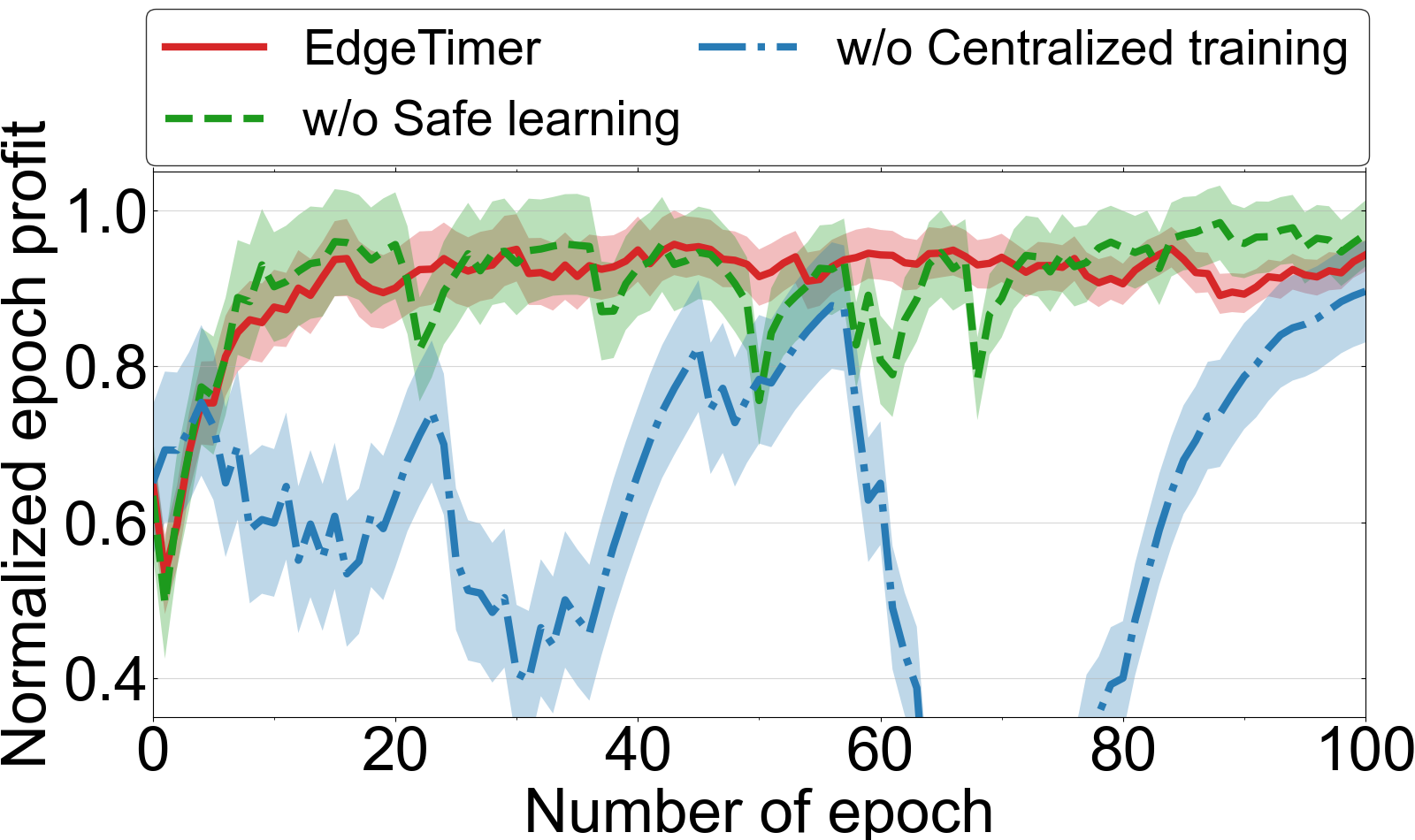} \\\footnotesize{(a) Offline profits.}
    \end{minipage}
    \begin{minipage}{0.49\linewidth} 
      \includegraphics[width=1\columnwidth]{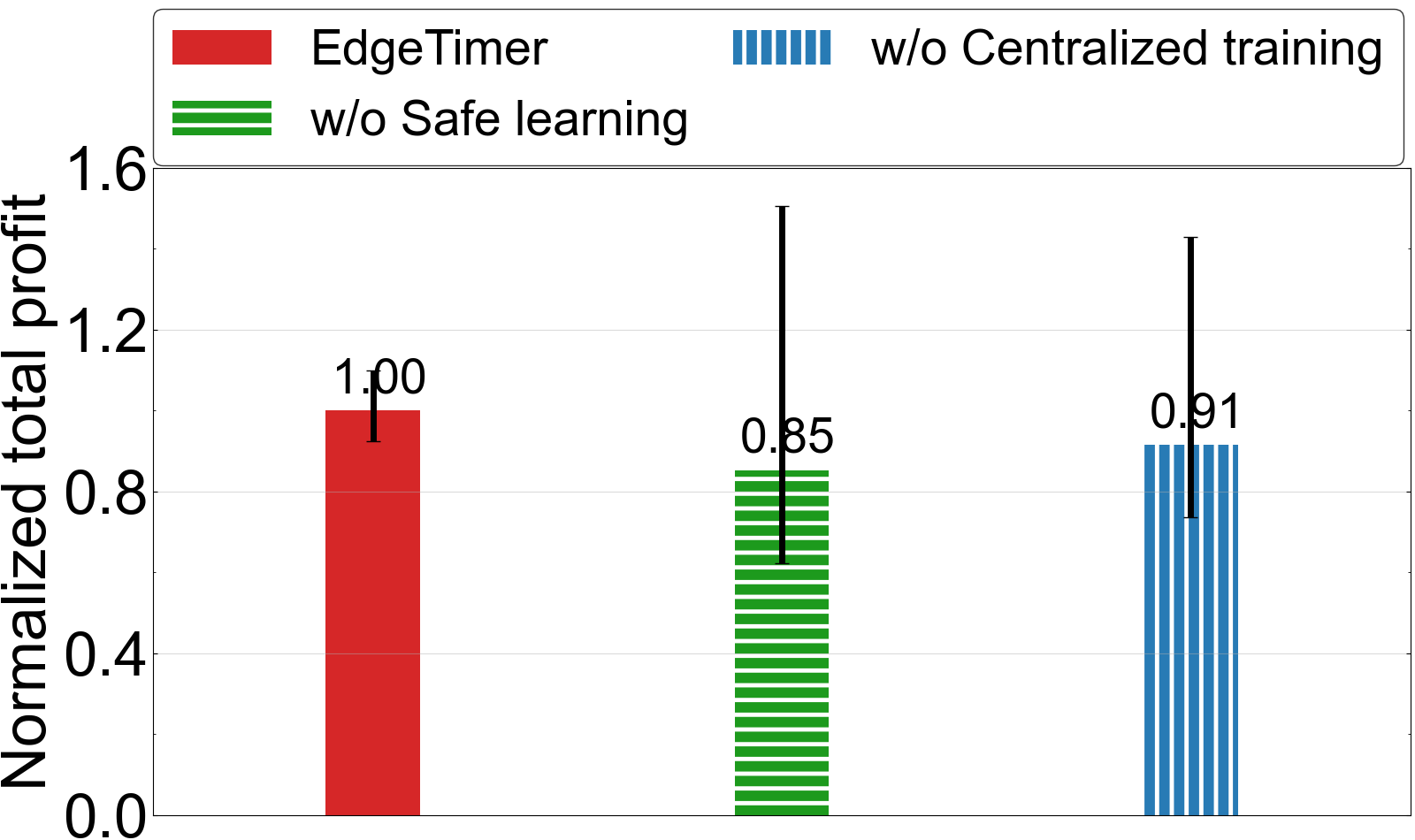} \\\footnotesize{(b) Online profits.}
    \end{minipage}
  \caption{Impact of the safe multi-agent DRL design on profit results.}
  \vspace{-1em}
  \label{Impact of safe}
\end{figure}
Each task is learned by the safe multi-agent DRL algorithm, where centralized training and safe learning are two key components to learn decentralized decisions under partial observations and ensure system reliability, respectively.
To verify their contributions, Fig. \ref{Impact of safe} compared the offline and online profit of EdgeTimer, EdgeTimer without centralized training, i.e, w/o Centralized training, and EdgeTimer without safe learning, i.e, w/o Safe learning.
EdgeTimer manages to reach a stable high profit than w/o Safe learning as it avoids high scheduling delay caused by unsafe actions. 
Furthermore, EdgeTimer overcomes the instability issues during the offline process encountered by w/o Centralized training, achieving $1.1\times$ higher online profit than that of w/o Centralized training.
This is because centralized training can utilize comprehensive information to evaluate policies of actor networks, providing a more precise direction for network updates.

\section{Related work}
\label{relatedwork}
\textbf{Profit Optimization in MEC.}
It is an important problem for service providers to maximize their profits that computed by deducting operation costs from user payments for allocated resources.
Many works are developed on reducing operation costs by modifying scheduling policies for service placement \cite{Adaptive2019, bifactor2021, online2019, modems2022}, task offloading \cite{guo2023efficient}, and resource allocation \cite{wang2018moera, jiao2017smoothed}.  
Another line of works aim to optimizing the price of resources based on uniform pricing \cite{Optimal2018, price2017, Pricing-Driven2021, Three2020}, differentiated pricing \cite{price2017, Pricing-Driven2021, Three2020}, and auction-based pricing strategies \cite{Mechanisms2022, TCDA2022, Three2020}, to benefit profits.
EdgeTimer opens a new dimension for profit maximization by optimizing and coordinating the update frequency of decisions in multi-layer scheduling, which enables to improve profits without changing the running scheduling algorithms of service providers.

\textbf{Multi-Timescale Scheduling in MEC.}
Multi-timescale scheduling is often applied to scheduling problems that contain multiple sequential decision-making for stability and cost saving.
Existing works adopt two timescales to update decisions for service placement and task offloading \cite{farhadi2019service, han2021tailored}, and task offloading and resource allocation \cite{liu2021learn}.
However, update timescales of different decisions in all the above works are fixed over time, synchronous at a fixed interval, and consistent for all edge servers.
EdgeTimer first provides an adaptive, asynchronous, and autonomous multi-timescale scheduling framework in edge computing.    

\textbf{DRL for Scheduling in MEC.}
DRL has been widely used in MEC to provide adaptive scheduling decisions for service placement \cite{han2021tailored}, task offloading \cite{han2022edgetuner, han2021tailored}, resource allocation \cite{xiong2020resource}, and server reservation \cite{zhang2021deepreserve}.
EdgeTimer integrates two tailored designs with the vanilla DRL algorithm to generate adaptive timescales for multi-layer heterogeneous scheduling decisions.
First, it adopts a three-layer hierarchical DRL framework to learn coupled policies of edge-cloud, edge-edge, and intra-edge scheduling for low complexity. 
Second, a safe multi-agent learning scheme is introduced in EdgeTimer to ensure reliability and accelerate learning.  

\section{Conclusion}
\label{conclusion}
In this paper, we study a new problem about how to update the scheduling decisions in edge computing to maximize the profit of service providers.
To this end, we propose EdgeTimer, an adaptive, asynchronous, and autonomous multi-timescale scheduling framework using deep reinforcement learning.
EdgeTimer enables each edge server to independently update respective decisions for edge-cloud, edge-edge, and intra-edge scheduling in a coordinated manner.
We design a three-layer hierarchical DRL to learn a hierarchy of policies with low complexity, and a safe multi-agent DRL for reliability enhancement.
Experiments on multi-pattern workloads demonstrate that EdgeTimer obtains up to $9.1\times$ higher profit than existing approaches and can support at least $45$ scheduling rules.
Service providers can use EdgeTimer as a plug-and-play approach by directly replacing scheduling rules in EdgeTimer with their own scheduling algorithms.
EdgeTimer opens a new dimension for profit improvement in edge computing.





\bibliographystyle{IEEEtran}

\end{document}